\documentclass[%
reprint,
superscriptaddress,
showpacs,preprintnumbers,
amsmath,amssymb,
aps,
prd,
longbibliography
]{revtex4-1}

\usepackage{float}
\usepackage{graphicx}
\usepackage{dcolumn}
\usepackage{bm}
\usepackage{bbold}
\usepackage{braket}
\usepackage{amssymb,amsmath}

\usepackage{color}
\usepackage{amsfonts}
\usepackage{subfigure}
\usepackage{array}

\newcommand{\Tr}{\ensuremath{\operatorname{Tr}}}

\newcolumntype{L}{>{\centering\arraybackslash}m{3cm}}

\definecolor{blue}{rgb}{0,0,1}

\definecolor{green}{rgb}{0,1,0}

\definecolor{red}{rgb}{1,0,0}

\definecolor{gray}{rgb}{.5,.5,.5}

\definecolor{darkgreen}{rgb}{.0,.5,.0}

\usepackage{xspace}
\usepackage{siunitx}
\usepackage{xfrac}
\usepackage{hyperref}
\usepackage[nameinlink]{cleveref}
\usepackage{appendix}
\usepackage{units}
\usepackage{xifthen}
\usepackage{xcolor}
\hypersetup{
	colorlinks,
	linkcolor={red!75!black},
	citecolor={blue!75!black},
	urlcolor={blue!75!black}
}
\def\Fig#1{\Cref{#1}} 
 
\def\fig#1{\Cref{#1}}

\def\Tab#1{\Cref{#1}}
\def\tab#1{\Cref{#1}}

\def\Eq#1{\Cref{#1}}
\def\eq#1{\Cref{#1}}
\def\eqref#1{\Cref{#1}}

\def\sec#1{\Cref{#1}}
\def\app#1{\hyperref[#1]{App.~\ref{#1}}}
\def\app#1{\Cref{#1}}

\def\lA0{{\langle A_0 \rangle}}
\def\bA0{{\bar{A}_0}}

\def\0#1#2{\frac{#1}{#2}}

\graphicspath{{./figures/}{./}}

\begin{document}
	
\title{Hyper-order baryon number fluctuations at finite temperature and density}

\author{Wei-jie Fu}
\affiliation{School of Physics, Dalian University of Technology, Dalian, 116024,
		P.R. China}
		
\author{Xiaofeng Luo}
\affiliation{Key Laboratory of Quark \& Lepton Physics (MOE) and Institute of Particle Physics,
		Central China Normal University, Wuhan 430079, China}

\author{Jan M. Pawlowski}
\affiliation{Institut f\"ur Theoretische Physik, Universit\"at Heidelberg, Philosophenweg 16, 69120 Heidelberg, Germany}
	\affiliation{ExtreMe Matter Institute EMMI, GSI, Planckstra{\ss}e 1, D-64291 Darmstadt, Germany}
	
\author{Fabian Rennecke}
\affiliation{Brookhaven National Laboratory, Upton, NY 11973, USA}
	
\author{Rui Wen}
\affiliation{School of Physics, Dalian University of Technology, Dalian, 116024,
		P.R. China}
	
\author{Shi Yin}
\affiliation{School of Physics, Dalian University of Technology, Dalian, 116024,
		P.R. China}

	
\begin{abstract}
Fluctuations of conserved charges are sensitive to the QCD phase transition and a possible critical endpoint in the phase diagram at finite density. In this work, we compute the baryon number fluctuations up to tenth order at finite temperature and density. This is done in a QCD-assisted effective theory that accurately captures the quantum- and in-medium effects of QCD at low energies. A direct computation at finite density allows us to assess the applicability of expansions around vanishing density. By using different freeze-out scenarios in heavy-ion collisions, we translate these results into baryon number fluctuations as a function of collision energy. We show that a non-monotonic energy dependence of baryon number fluctuations can arise in the non-critical crossover region of the phase diagram. Our results compare well with recent experimental measurements of the kurtosis and the sixth-order cumulant of the net-proton distribution from the STAR collaboration. They indicate that the experimentally observed non-monotonic energy dependence of fourth-order net-proton fluctuations is highly non-trivial. It could be an experimental signature of an increasingly sharp chiral crossover and may indicate a QCD critical point. The physics implications and necessary upgrades of our analysis are discussed in detail.
\end{abstract}

\maketitle

\section{Introduction}\label{sec:int}

Some of the most challenging questions of heavy-ion physics are related to the transition from the early, non-equilibrium, state of quarks and gluons to the final hadronic states after chemical freeze out, which is observed in experiments. Unravelling this dynamics necessitates a thorough grasp on the physics in the QCD phase structure close to the confinement-deconfinement and chiral transitions. This regime is strongly correlated with highly non-trivial dynamics. Understanding this part of the phase structure, including the location and dynamics of a potential critical end point (CEP), plays a pivotal role in understanding phases of strongly interacting nuclear matter under extreme conditions. For works on the phase structure of QCD, covering experiment and theory see, e.g.,  \cite{Stephanov:2007fk, Friman:2011zz, NICA-white, Luo:2017faz, Dainese:2019xrz, Bzdak:2019pkr, Fischer:2018sdj, Fu:2019hdw, Bazavov:2020bjn, Borsanyi:2020fev,Lu:2016htm,Yang:2013yeb,Sako:2019hzh}, where theory covers first principles functional approaches and lattice simulations.

Fluctuations of conserved charges are very sensitive to the physics of the strongly correlated regime that governs the transition from the quark-gluon plasma (QGP) to the hadronic phase.
They provide detailed information on the underlying dynamics. This includes, but is not limited to, possible experimental signatures of a CEP \cite{Luo:2017faz}. It has for example been proposed in \cite{Stephanov:1999zu, Stephanov:2008qz, Stephanov:2011pb} that non-monotonic variations of conserved charge fluctuations as functions of the beam energy can arise from critical physics in the vicinity of a CEP. During the last decade, significant fluctuation measurements have been performed in the first phase of the Beam Energy Scan (BES-I) program at the Relativistic Heavy Ion Collider (RHIC), involving various cumulants of net-proton, net-charge and net-kaon multiplicity distributions \cite{Adamczyk:2013dal, Adamczyk:2014fia, Luo:2015ewa, Adamczyk:2017wsl, Adam:2019xmk}.
Remarkably, very recently the STAR collaboration has reported the first evidence of a non-monotonic variation in the kurtosis (multiplied by the variance) of the net-proton number distribution as a function of the collision energy with $3.1\,\sigma$ significance for central collisions \cite{Adam:2020unf}. The measurements have been extended to the sixth-order cumulants of net-proton and net-charge distributions, for preliminary results from STAR see  \cite{Nonaka:2020crv, Pandav:2020uzx}.

Recent first-principle QCD calculations at finite temperature and density, within both the functional renormalisation group (fRG) and Dyson-Schwinger equations (DSE), show that the transition from the QGP to the hadronic phase is a crossover which becomes sharper with increasing baryon chemical potential, $\mu_B$, for $\mu_B/T \lesssim 4$ \cite{Fischer:2018sdj, Fu:2019hdw, Braun:2019aow, Isserstedt:2019pgx, Gao:2020qsj, Gao:2020fbl}. Beyond this region, a CEP might occur, but quantitative reliability of the theory computations cannot be guaranteed within the present approximations \cite{Fischer:2018sdj, Fu:2019hdw, Braun:2019aow, Gao:2020fbl}. In addition, critical physics may only be observable in a very small region around the CEP, see e.g.\ \cite{Schaefer:2006ds}. 
Since the available RHIC data is limited to $\mu_B/T\lesssim 3$, it is important to understand how conserved charge fluctuations are affected by the increasingly sharp crossover away from a regime with critical scaling.

To address these open questions related to the physics of strong correlations in the QCD phase diagram, we study in detail the $T$- and $\mu_B$-dependence of net-baryon number fluctuations in the range $\mu_B/T \lesssim 3$. We present results for fluctuations up to tenth order (where we refer to everything above fourth order as \emph{hyper-order}), including comparisons to available results from RHIC \cite{Adam:2020unf, Nonaka:2020crv, Pandav:2020uzx} and predictions for the beam energy dependence of fluctuations where no experimental results are available yet.

To facilitate the comparison between theory and experiment, $T$ and $\mu_B$ can be mapped onto the beam energy per nucleon, $\sqrt{s_{\rm NN}}$, via phenomenological freeze-out curves \cite{BraunMunzinger:2003zd}. While these curves are expected to be close to the QCD crossover at large beam energies (corresponding to small $\mu_B$) \cite{BraunMunzinger:2003zz}, they may move away from the transition region at low energies (i.e.\ larger $\mu_B$) \cite{Floerchinger:2012xd}. This can affect the beam energy dependence of particle number fluctuations, and requires a detailed understanding of the physics also outside the critical region.

Aside from their phenomenological relevance, net-baryon number fluctuations at finite $\mu_B$ can also be used to assess the reliability of extrapolations of thermodynamic quantities to finite $\mu_B$ based on a Taylor expansion at $\mu_B = 0$. Such a strategy is commonly used in lattice QCD simulations, where a sign problem prevents direct simulations at finite $\mu_B$, see e.g.\ \cite{Bazavov:2012vg, Borsanyi:2013hza, Borsanyi:2014ewa, Bazavov:2017dus, Bazavov:2017tot, Borsanyi:2018grb, Bazavov:2020bjn}. By comparing the results of direct computations at finite $\mu_B$ to the ones obtained from extrapolations at $\mu_B=0$, we study the range of validity of a Taylor expansion at a given order self-consistently. Understanding the limitations of such an extrapolation is also relevant for phenomenologically constructed equations of state, as, e.g., in \cite{Parotto:2018pwx}, where the non-critical physics at finite $\mu_B$ crucially rely on this extrapolation.

All this is addressed within a QCD-assisted low-energy effective field theory (LEFT) which is described in detail in the next section. We use first-principles QCD-results on the $T$-dependence of the kurtosis and the $\mu_B$-dependence of the chiral phase boundary to map the in-medium scales of the LEFT onto QCD. This improves the reliability of our predictions, in particular at finite $\mu_B$. Non-perturbative quantum-, thermal- and density fluctuations are taken into account with the functional renormalisation group (fRG). This work therefore is a significant upgrade of previous work in \cite{Fu:2015naa, Fu:2015amv, Fu:2016tey}, where net-baryon number fluctuations up to fourth order have been studied. The present QCD-assisted LEFT approach has various advantages. Most importantly, it is directly embedded in QCD as the relevant low-energy degrees of freedom emerge dynamically from systematically integrating-out the fast partonic modes of QCD \cite{Mitter:2014wpa, Braun:2014ata, Rennecke:2015eba, Cyrol:2017ewj, Fu:2019hdw}. In addition, this approach allows us to capture both critical and non-critical effects in the QCD phase diagram. This entails in particular that our results agree with the results of lattice QCD at small $\mu_B$ and show the correct universal behaviour of QCD in the vicinity of the CEP, i.e.\ $3d$ Ising universality.

Concerning the existence and location of the latter we add that the first-principles results in \cite{Fischer:2018sdj, Fu:2019hdw, Braun:2019aow, Isserstedt:2019pgx, Gao:2020qsj, Gao:2020fbl} include a CEP
in a region of $450\,\mathrm{MeV} \lesssim\mu_B\lesssim 650\,\mathrm{MeV}$ and therefore outside the regime of quantitative reliability of these computations.
This suggests that the experimental detection of a CEP requires explorations of the high-$\mu_B$ in the region with $\mu_B/T_c\gtrsim 4$. 
Moreover, the direct experimental measurement of the CEP may be very challenging as it requires very high statistics, and predictions that signal critical dynamics can be further complicated by non-equilibrium effects. 
In the present work we shall therefore also outline how the location of a CEP could be constrained based on data in the crossover region, without the necessity of observing critical scaling. On the theoretical side this asks for first principles QCD studies for $\mu_B/T\gtrsim 4$. In turn, a first experimental step towards this goal is the solidification of experimental observation of the non-monotonic energy dependence of fourth-order net-proton fluctuations. Both is safely beyond the scope of the present work. 

This paper is organised as follows: In \sec{sec:FRG} we give a brief introduction to the fRG-approach to QCD and low-energy effective theories, including their mutual relationship. Thermodynamics and the hyper-order baryon number fluctuations are discussed in \sec{sec:hyper-fluc}. In \sec{sec:num}, we first introduce a systematic scale-matching procedure between QCD and the low-energy effective theory. We then present our numerical results and compare them to lattice QCD simulations and experimental measurements. A summary with conclusions is given in \sec{sec:summary}. Technical details regarding the flow equations are presented in the appendices.

	
\section{QCD and emergent low-energy effective theories}
\label{sec:FRG}
	
%
\begin{figure}[t]
\includegraphics[width=0.45\textwidth]{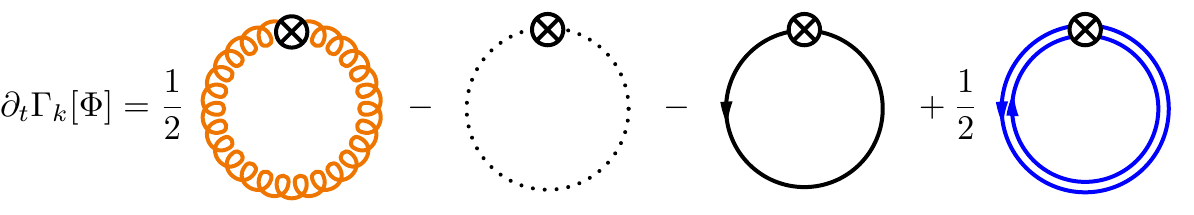}
\caption{Diagrammatic representation of the QCD flow equation. The lines stand for the full propagators of gluon, ghost, quark, and mesons, respectively. The arrows in quark and meson lines indicate the quark number (baryon number) flow. The crossed circles represent the infrared regulators.}\label{fig:QCD_equation}
\end{figure}
%

At low momentum scales the quark-gluon dynamics of QCD successively decouple due to the QCD mass gap and spontaneous chiral symmetry breaking. This decoupling also applies to most dynamical (hadronic) low energy degrees of freedom at even lower energies, finally leaving us with dynamical pions and hence with chiral perturbation theory. Indeed, this successive decoupling is at the root of the success of chiral effective field theory. 
	
The functional renormalisation group approach to QCD with its successive integrating-out of momentum modes is ideally suited to follow and study this decoupling. Diagrammatically, this is already seen within the flow equation for the QCD effective action, depicted in \fig{fig:QCD_equation}. The different lines stand for the full non-perturbative propagators of gluons, ghosts, quarks and emergent low-energy degrees of freedom (hadrons in our case), where the loop momentum $q$ is restricted by the infrared cutoff scale $k$, $q^2\lesssim k^2$. In this setup, emergent bound states can be incorporated systematically by dynamical hadronisation \cite{Gies:2001nw, Gies:2002hq, Pawlowski:2005xe, Floerchinger:2009uf}. For quantitative QCD applications in the vacuum see \cite{Braun:2014ata, Rennecke:2015eba, Mitter:2014wpa, Cyrol:2017ewj}, for further conceptual developments and the application to the QCD phase structure important for the present work see \cite{Fu:2019hdw}. The decoupling is apparent in this framework as the propagators carry the mass gaps $m_\textrm{gap}$ of gluons and quarks and for cutoff scales $k\ll m_\textrm{gap}$ of a given field the respective loop tends towards zero. 
	
More importantly, in this way the emergent low-energy effective theory is naturally embedded in QCD, and its ultraviolet parameters (at $\Lambda\lesssim1$\,GeV) as well as further input may be directly computed from QCD, leading to  \textit{QCD-assisted} low-energy effective theories. We emphasise that this procedure does not lead to a unique LEFT. The dynamical degrees of freedom of QCD-assisted LEFTs at $\Lambda\lesssim1$\,GeV depend on the dynamical hadronisation procedure applied within QCD-flows. This setup and the QCD-embedding entail that, provided the relevant quantum, thermal and density fluctuations of low energy QCD are taken into account in the QCD-assisted LEFT at hand, all QCD-assisted LEFTs encode the same physics, namely that of low energy QCD.

This leads to an equivalence relation of QCD-assisted Polyakov-loop--enhanced NJL-type LEFTs (PNJL), Polyakov-loop--enhanced QM LEFTs (PQM) and variations including higher meson multiplets and/or diquarks and baryons. We emphasise again that this equivalence relation only holds if low energy quantum, thermal and density fluctuations are taken into account. For more details see in particular \cite{Fu:2019hdw}, and the recent review \cite{Dupuis:2020fhh}. Most prominently this embedding has been used for determining the temperature-dependence of the Polyakov loop potential, see \cite{Haas:2013qwp, Herbst:2013ufa}. This setup was then applied to the computation of fluctuations in \cite{Fu:2015amv, Fu:2015naa, Fu:2016tey, Wen:2018nkn, Yin:2019ebz}. 
	
In summary this entails, that for sufficiently small momenta $k$, temperatures $T$, and also density or quark chemical potential $\mu_q$, the gluon (and ghost) loop in \fig{fig:QCD_equation} decouple from the dynamics, and only provide a non-trivial glue background at finite temperature and chemical potential. The latter is taken into account with the Polyakov loop potential discussed in detail below. 
	
In the present work, we build upon previous investigations  of the skewness and kurtosis of baryon number distributions \cite{Fu:2015naa, Fu:2015amv, Fu:2016tey}, and baryon-strangeness correlations \cite{Fu:2018qsk, Fu:2018swz} within QCD-assisted LEFTs with the fRG. The present LEFT is an upgrade of those used in the works above, and includes the quantum, thermal and density dynamics of quarks, pions and the sigma mode in a Polyakov loop background. It is a QCD-assisted PQM. As argued above, for low enough chemical potential, this model sufficiently close to QCD, and leads to results that are independent of the LEFT at hand.

For further investigations of fluctuation observables within the fRG-approach to low-energy effective theories see e.g.\  \cite{Skokov:2010wb, Skokov:2010uh, Friman:2011pf, Morita:2014fda, Almasi:2017bhq}, the Dyson-Schwinger approach has been used in e.g.\  \cite{Xin:2014ela, Isserstedt:2019pgx},  for mean-field investigations see e.g.\ \cite{Fu:2009wy, Fu:2010ay, Karsch:2010hm, Schaefer:2011ex, Li:2018ygx}.  These functional works can be further adjusted and benchmarked with results from lattice QCD simulations  \cite{Bazavov:2012vg, Borsanyi:2013hza, Borsanyi:2014ewa, Bazavov:2017dus, Bazavov:2017tot, Borsanyi:2018grb, Bazavov:2020bjn}, at high temperatures, $T\gtrsim T_c$, and vanishing $\mu_B$. In turn, at finite $\mu_B$, and in particular for $\mu_B/T_c\gtrsim 3$, lattice simulations are obstructed by the sign problem.


\subsection{2-flavour setup}\label{sec:Nf2LEFT}

For the physics of fluctuations we are interested in scales below approximately $1$\,GeV. We restrict ourselves to $k\lesssim 700$\,MeV and temperatures and quark chemical potentials $T,\mu_q \lesssim 200$\,MeV. In this regime the only relevant quarks are the light quarks $q=(u,d)$ and the strange quark $s$. The latter, while changing the momentum-scale running of the correlation functions, has subleading effects on the form of the fluctuations. Hence, the effect of the momentum-scale running induced by strange fluctuations will be mimicked here by an appropriate scale-matching detailed in \sec{subsec:scale}. 

We also include the lowest lying hadronic resonances, the pion ${\bm\pi}=(\pi^\pm,\pi^0)$, and, for symmetry reasons, the scalar resonance $\sigma$ as effective low energy degrees of freedom. Within QCD flows these fields are emergent low energy degrees of freedom at cutoff scales $k\simeq 1$\,GeV, that are taken care of with dynamical hadronisation in e.g.\ \cite{Fu:2019hdw}. At the present low energy scales $k\leq 700$\,MeV, they are fully dynamical, and hence are part of the effective action at the initial cutoff scale. The other members of the lowest lying multiplet as well as further hadronic resonances produce rather subleading contributions to the offshell dynamics and hence are dropped. The mesonic fields are stored in an O(4) scalar field $\phi=(\sigma, {\bm\pi})$ with the corresponding chiral invariant $\rho = \phi^2/2$. 

Quantum, thermal and density fluctuations with scales $k\lesssim \Lambda= 700$\,MeV are taken into account within the fRG, whose dynamics is now reduced to the last two loops in \fig{fig:QCD_equation}. The respective effective action of QCD in the low energy regime is approximated by 
\begin{align}
\Gamma_k=&\int_x \bigg\{Z_{q,k}\bar{q} \Big [\gamma_\mu \partial_\mu -\gamma_0(\mu+i g A_0) \Big ]q+\frac{1}{2}Z_{\phi,k}(\partial_\mu \phi)^2 \nonumber\\[2ex]
&\hspace{.5cm}+h_k\,\bar{q}\,
\left(\tau^0\sigma+\bm\tau
\cdot\bm{\pi}\right)\,q +V_k(\rho,A_0)-c\sigma \bigg\}\,,\label{eq:action}
\end{align}
with $\int_{x}=\int_0^{1/T}d x_0 \int d^3 x$ and $\tau=1/2 (\mathbb{1}, i \gamma_5 \bm \sigma)$. We assume isospin symmetry and the corresponding chemical potential flavour-matrix is given by $\mu = \text{diag}(\mu_q,\mu_q) = \frac{1}{3}\text{diag}(\mu_B,\mu_B)$. $Z_{q,k}$ and $Z_{\phi,k}$ are the wave function renormalisations for the light quarks and the meson respectively. Further running couplings considered are the Yukawa coupling $h_k$, the scattering between quarks and mesons, as well as the effective potential $V_k(\rho,A_0)$, that describes the multi-scattering of mesons in the non-trivial glue background present at finite temperature and chemical potential.
	
The flow equation for the effective action \eq{eq:action}, and that for $V_k, h_k, Z_{\phi,q}$ is described in \app{app:fRG} and \app{app:flowV}. 
The initial condition for $\Gamma_k$ at the initial cutoff scale $k=700$\,MeV is described in \app{app:Ini}. 
	
The potential $V_k(\rho, A_0)$ has contributions $V_{\mathrm{glue},k}(A_0)$ from offshell glue fluctuations (first two diagrams in \fig{fig:QCD_equation}), and contributions $V_{\mathrm{mat},k}(\rho,A_0)$ from the quark loop (third diagram in \fig{fig:QCD_equation}). This leads us to 
\begin{align}
V_k(\rho,A_0)=&V_{\mathrm{glue},k}(A_0)+V_{\mathrm{mat},k}(\rho,A_0)\,,\label{eq:Vtotal}
\end{align}
The first contribution is typically reformulated in terms of the Polyakov loop $L(A_0)$, while the latter is directly computed from the present low energy flow. This allows us to trade the $A_0$-dependence for that of the traced Polyakov loop, $L(A_0), \bar L(A_0)$, see \app{app:gluepot}, \Eq{eq:Lloop}, leading us to to the final form of our potential, 
\begin{align}
V_k(\rho, L,\bar L)= V_k(\rho, A_0)\,. 
\end{align}
More details can be found in \app{app:gluepot}. 

In conclusion, the QCD-assisted LEFT described above and used in the present work, is a PQM-type model, e.g.\  \cite{Schaefer:2007pw, Schaefer:2009ui, Skokov:2010wb, Herbst:2010rf, Skokov:2010uh, Karsch:2010hm, Schaefer:2011ex, Morita:2011jva, Skokov:2011rq, Mintz:2012mz, Haas:2013qwp, Herbst:2013ail, Herbst:2013ufa, Fu:2015amv, Fu:2015naa, Fu:2016tey, Stiele:2016cfs, Sun:2018ozp, Fu:2018qsk, Fu:2018swz, Wen:2018nkn, Yin:2019ebz, Hansen:2019lnf}. Quantum, thermal and density fluctuations below $\Lambda=700$\,MeV are taken account with the functional renormalisation group, and the setup is well-embedded in functional QCD. As argued above, within the present, and analogous, elaborate approximation, the respective results (for fluctuation observables) for all QCD-assisted LEFTs match those of QCD for sufficiently low density. Therefore, we will refer to this model from now on as generic QCD-assisted LEFT.


\subsection{$2+1$-favour scale-matching in $2$-flavour QCD}\label{subsec:scale}
	
The current QCD-assisted LEFT setup enables us to compute thermodynamic observables and in particular hyper-order baryon number fluctuations. However, as already briefly discussed in \sec{sec:Nf2LEFT}, we have dropped the dynamics of the strange quark. While we expect sub-dominant effects on hyper-fluctuations, the $s$-quark influences the momentum running of the correlations in the ultraviolet. 
	
Importantly, in \cite{Fu:2019hdw} it has been observed on the basis of genuine $N_f=2$ and $N_f=2+1$ flavour computations in QCD, that the latter effect is well approximated by a respective universal scale-matching of the 2-flavour results even in QCD. Such a scale-matching has already led to a quantitative agreement of thermodynamics and kurtosis within the current LEFT setup with lattice results, see \cite{Fu:2015amv, Fu:2015naa, Fu:2016tey}. Thus, the present scale-matching entails that we use information on the $T$- and $\mu_B$-dependence of well-determined quantities in QCD. This leads to an improved reliability of our results of finite $T$ and $\mu_B$, as in-medium effects in the QCD-assisted LEFT are directly connected to in-medium effects in QCD.


\subsubsection{2- to 2+1-flavour scale-matching in QCD}

Given its relevance for the predictive power of the present LEFT within a QCD scale-matching procedure we briefly recall the respective results in \cite{Fu:2019hdw}: There, the phase boundaries of 2- and 2+1-flavour QCD have been computed within the fRG approach. These results allow us to evaluate the reliability of even linear scale-matchings of temperatures and chemical potentials in 2- and 2+1-flavour QCD introduced by 
\begin{align}\nonumber 
T^{(N_f=2)} &=c_{_{T}}\,T^{(N_f=2+1)} \,,\\[1ex]
\mu_B^{(N_f=2)} &= c_{\mu_B}\,\mu_{B}^{(N_f=2+1)} \,.
\label{eq:rescale}\end{align}
With such a linear scale-matching the scaling factors $c_{_{T}}, c_{\mu_B}$ can be determined by evaluating the relations at a specific temperature and chemical potential. 

For the thermal scale-matching we naturally take $(T,\mu_B)=(T_c,0)$, the crossover temperature at vanishing chemical potential. In \cite{Fu:2019hdw} the crossover temperatures have been determined with thermal susceptibilities of the renormalised light chiral condensate. Then, the linear rescaling of the 2-flavour chiral crossover temperature to the 2+1-flavour crossover temperature is done with 
\begin{align}\label{eq:cTQCD}
T_c^{(N_f=2)} = c_T^{\textrm{QCD}}\, T_c^{(N_f=2+1)}\,, \qquad  c_T^{\textrm{QCD}}=1.1\,. 
\end{align} 
For the matching of the chemical potentials we use the curvature $\kappa$ of the phase boundary at vanishing $\mu_B=0$, 
\begin{align}
\frac{T_c(\mu_B)}{T_c}&=1-\kappa \left(\frac{\mu_B}{T_c}\right)^2+\lambda \left(\frac{\mu_B}{T_c}\right)^4+\cdots\,.\label{eq:curv}
\end{align}
Adjusting the 2-flavour curvature $-\kappa\, \mu_B^2/T_c^2$ to the 2+1-flavour one leads us to the relation 
\begin{align}
c^\textrm{QCD}_{\mu_B}&=c^{\textrm{QCD}}_{_{T}}\left(\frac{\kappa^{(N_f=2+1)}}{\kappa^{(N_f=2)}}\right)^{1/2}\,, \qquad c_{\mu_B}^\textrm{QCD}=0.99 \,.\label{eq:cmuQCD}
\end{align}
The value $c_{\mu_B}^\textrm{QCD}\approx 1$ entails that the change in the curvature coefficient $\kappa$ is balanced by that of the temperature.  

%
\begin{figure}[t]
\includegraphics[width=0.45\textwidth]{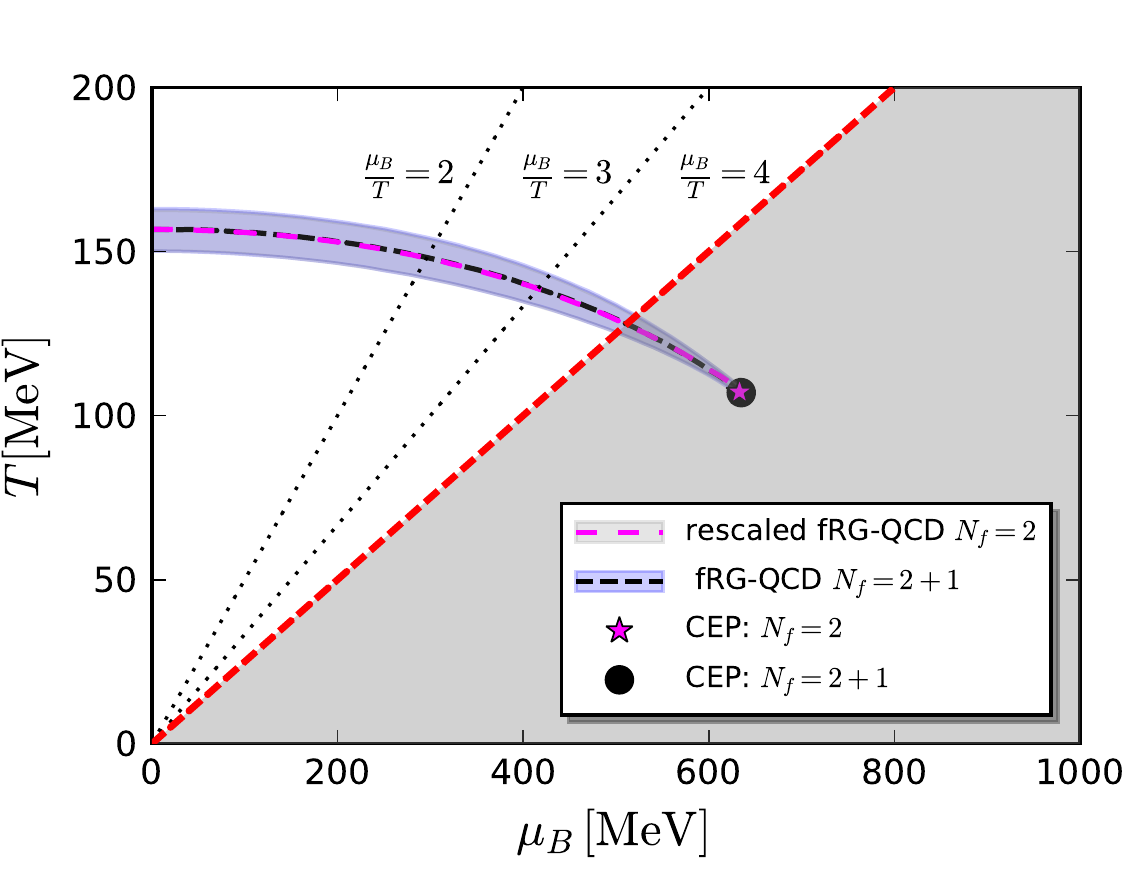}
\caption{Phase boundaries of 2- and 2+1-flavour QCD from \cite{Fu:2019hdw} with a 2+1-flavour scale-matching of the 2-flavour data at the crossover temperature and $\mu_B=0$. The bands denote the width of the chiral crossover. The scale-matched 2-flavour phase boundary agrees quantitatively with the genuine 2+1-flavour one including the location of the critical end point. The dashed line at $\mu_B/T=4$ constitutes the reliability bound of the computations in \cite{Fu:2019hdw} based on the potential emergence of new degrees of freedom discussed in \cite{Fu:2019hdw, Braun:2019aow, Fischer:2018sdj}. The dashed lines at $\mu_B/T=2,3$ are reliability estimates of lattice results as well as old ones from functional approaches.  }\label{fig:QCD-scalematching}
\end{figure}
%
The fourth-order expansion coefficient $\lambda$ is found to be very small in both functional, \cite{Fu:2019hdw, Gao:2020fbl, Gao:2020qsj} as well as lattice computations, \cite{Bazavov:2018mes,Borsanyi:2020fev}. Moreover, the results for the phase boundary at finite chemical potential in \cite{Fu:2019hdw, Gao:2020fbl, Gao:2020qsj, Fischer:2018sdj} reveal that the phase boundary is still described well by the leading  order expansion with $\mu_B^2$-terms. We estimate, that this prediction is quantitatively reliable within $\mu_B/T\lesssim 4$, using results from \cite{Fu:2019hdw, Gao:2020fbl, Gao:2020qsj, Fischer:2018sdj, Braun:2019aow}. This covers the regime studied in the present work. 

Applying the two scale-matching relations in \Eq{eq:rescale} with the coefficients \eq{eq:cTQCD} and \eq{eq:cmuQCD} to the 2- and 2+1-flavour data of the QCD phase boundary in \cite{Fu:2019hdw} leads us to \fig{fig:QCD-scalematching}. In conclusion, this impressive agreement provides non-trivial support for the scale-matching procedure in QCD.


\subsubsection{2- to 2+1-flavour scale-matching in LEFTs}	\label{sec:scaleLEFT}

The convincing quantitative accuracy of the linear scale-matching analysis presented for QCD in the last section also sustains its use in the LEFT within the present work. Note however, that we cannot simply take over the above QCD-relations for the present LEFT, which lacks the backcoupling of the glue-dynamics on both large temperature and chemical potential physics. Still, the dominance of the leading order term $-\kappa \mu_B^2/T_c^2$ in the model reflects the same property in QCD. This allows us to employ a respective linear scale-matching for $\mu_B/T\lesssim 4$ as studied in the present work.

%
\begin{figure}[t]
\includegraphics[width=0.45\textwidth]{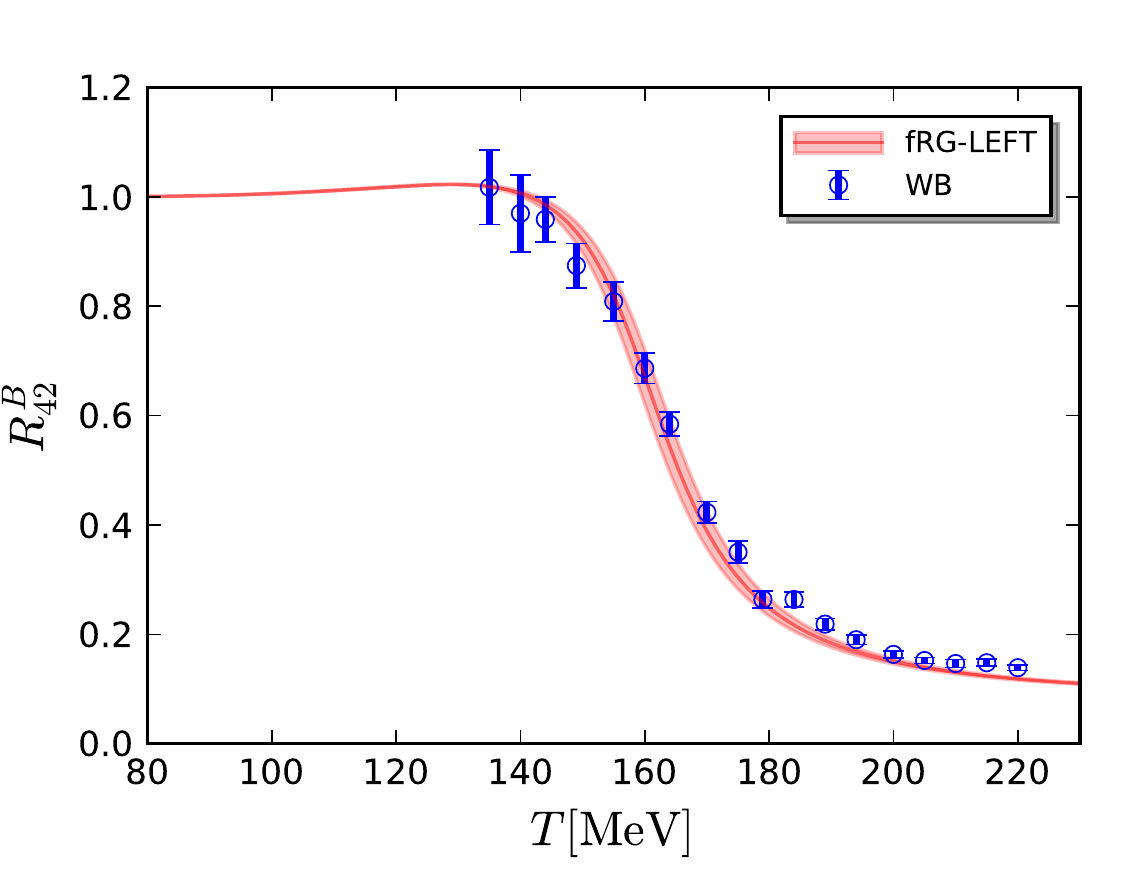}
\caption{Matching of the temperature-scale in the QCD-assisted 2-flavour LEFT with $R_{42}^B$ in \Eq{eq:Rnm}, using the 2+1-flavour lattice results of \cite{Borsanyi:2018grb}. This leads to $c_{_{T}}=1.247(12)$ in \Eq{eq:cTQCD}. The $T/T_c$-dependence of $R_{42}^B$ is a prediction of the QCD-assisted LEFT, and agrees quantitatively with the lattice results.}\label{fig:T-adjust}
\end{figure}
%
Analogously to QCD we choose the chiral crossover temperature at vanishing chemical potential, $(T,\mu_B)=(T_c,0)$ for fixing the scale factor $c_{_T}$. Moreover, in the present work we are interested in fluctuations of conserved charges. Hence, instead of the renormalised condensate we use the kurtosis of baryon number fluctuations, or rather $R_{42}^B=\chi_4^B/\chi_2^B$, for the definition see \eq{eq:suscept} and \eq{eq:Rnm} with \eq{eq:chiB2}, \eq{eq:chiB4}. This leads us to the following determination of $c_T$: While the temperature-dependence of $R^{B}_{42}$ is a prediction of the LEFT, its absolute temperature has to be adjusted. This is done by minimising the $\chi^2$ of the difference between the lattice result and the LEFT-prediction as a function of the rescaled absolute temperature $c_T T_c$, leading us to 
\begin{align}
c_{_{T}}=1.247(12)\,,\label{eq:c_T-LEFT}
\end{align}
The respective result for $R^B_{42}$ is shown in \Fig{fig:T-adjust} in comparison to the lattice result from \cite{Borsanyi:2018grb}. The two curves match quantitatively supporting the predictive power of the LEFT. 
	
For the scale-matching of $\mu_B$ with the curvature $-\kappa \mu_B^2/T_c^2$ we have a plethora of results from state of the art functional approaches: $\kappa=0.0142(2)$ in  \cite{Fu:2019hdw}, $\kappa=0.0150(7)$ in \cite{Gao:2020qsj} and $\kappa=0.0147(5)$ in \cite{Gao:2020fbl}, the very recent update of \cite{Gao:2020qsj}. Lattice results are provided with $\kappa=0.015(4)$ in \cite{Bazavov:2018mes}, $\kappa=0.0149(21)$ in \cite{Bellwied:2015rza}, $\kappa=0.0153(18)$ in \cite{Borsanyi:2020fev}. Both, functional and lattice results agree within the respective (statistical and systematic) errors with $\kappa\approx 0.015$. 

Having adjusted the temperature with results from the WB-collaboration, \cite{Bellwied:2015rza}, we use $\kappa=0.0153(18)$ from \cite{Borsanyi:2020fev} for internal consistency. Note that the results presented here do only change marginally if using one $\kappa$ in the range $\kappa=(0.0142 - 0.0153)$. Within the current LEFT we obtain $\kappa_{_\textrm{LEFT}}=0.0193$. In comparison, $\kappa_{_\textrm{LEFT}}$ is larger than the 2-flavour QCD result in \cite{Fu:2019hdw} with $\kappa=0.0179(8)$. This reflects the lack of glue-dynamics in the LEFT. We use this in the relation \eq{eq:cmuQCD} instead of $\kappa^{(N_f=2)}$, and arrive at  
\begin{align}
c_{\mu_B}=c_{_{T}}\left(\frac{\kappa^{N_f=(2+1)}}{\kappa_{_{\textrm{LEFT}}}}\right)^{1/2}=1.110(66)\,, \label{eq:cmu-LEFT}
\end{align}
with the LEFT-$c_{_T}$ from \Eq{eq:c_T-LEFT}. 
	
In summary, as our first step towards a quantitative prediction for hyper-order baryon number fluctuations, in this work we do not deal with the strange quark as a dynamical degree of freedom for the moment, but rather take into account its effect on the modification of the momentum-scale running via an appropriate scale-matching as shown in \Eq{eq:rescale}. The validity of the scale-matching relations between 2- and 2+1-flavour QCD has been well verified in this section by means of results from the first-principle functional QCD in \cite{Fu:2019hdw}. These relations were applied to the present QCD-assisted LEFT. The scale-matching was done with two observables relevant for the fluctuation physics studied here: $R^{B}_{42}$ as a function of $T$ and the curvature of the phase boundary $\kappa$, both at vanishing chemical potential. This led us to the coefficients \eq{eq:c_T-LEFT} and \eq{eq:cmu-LEFT} in \Eq{eq:rescale}. The errors in these coefficients determine the errors of our results in \sec{sec:num}.

	
\section{Thermodynamics and Hyper-order baryon number fluctuations}
\label{sec:hyper-fluc}

The thermodynamic potential in the LEFT at finite temperature and baryon chemical potential is readily obtained from the effective action in \Eq{eq:action}, or rather from its integrated flow: we evaluate the effective action on the solution of the quantum equations of motion (EoMs). In the present work we consider only homogeneous (constant) solutions,  $(\sigma_\textrm{EoM}, A_{0,\textrm{EoM}})$ with
\begin{align}\label{eq:EoMs}
\frac{\partial V(\rho, L,\bar L)}{\partial \sigma} =   \frac{\partial V(\rho, L,\bar L)}{\partial L} = \frac{\partial V(\rho, L,\bar L)}{\partial \bar L} =0\,, 
\end{align}
while the quark fields vanish on the EoMs, $q,\bar q=0$. We also note that the assumption of homogeneous solutions has to be taken with a grain of salt for larger chemical potentials with $\mu_B/T\gtrsim 4$, see \cite{Fu:2019hdw}. Such a scenario has been investigated in LEFTs, see e.g.\ the review \cite{Buballa:2014tba} and references therein. 

With these preparations we are led to the grand potential $\Omega[T,\mu_B]= V_{k=0}(\rho, L,\bar L)$, the effective potential, evaluated at vanishing cutoff scale $k=0$. It reads   
\begin{align}
\Omega[T,\mu_B]=&V_{\mathrm{glue}}(L, \bar L)+V_{\mathrm{mat}}(\rho, L, \bar L)-c\sigma\,,\label{eq:Omega}
\end{align}
where the gluonic background field $A_0$ in \Eq{eq:Vtotal} has been reformulated in terms of the Polyakov loop $L$ and its complex conjugate $\bar L$. As mentioned before, the matter sector of the effective potential is integrated out towards the IR limit $k=0$, for details see \app{app:flowV}. In turn, the glue sector is independent of $k$, see \app{app:gluepot}. The pressure of the system follows directly from the thermodynamic potential,
\begin{align}
p=&-\Omega[T,\mu_B]\,.\label{eq:pres}
\end{align}
The generalised susceptibilities of the baryon number $\chi^B_n$ are defined through the $n$-th order derivatives of the pressure w.r.t.\ the baryon chemical potential, to wit,
\begin{align}
\chi_n^{B}&=\frac{\partial^n}{\partial (\mu_B/T)^n}\frac{p}{T^4}\,.\label{eq:suscept}
\end{align}
To remove the explicit volume dependence, it is advantageous to consider the ratio between the $n$- and $m$-th order susceptibilities, defined by,   
\begin{align}
R_{nm}^{B}&=\frac{\chi_n^{B}}{\chi_m^{B}}\,.
\label{eq:Rnm}
\end{align}
The generalised susceptibilities are related to various cumulants of the baryon number distribution, which can be measured in heavy-ion collision experiments through the cumulants of its proxy, i.e., the net proton distribution, see, e.g., \cite{Luo:2017faz} for details. For the lowest four orders we get, 
\begin{align}
\chi^B_1=&\frac{1}{VT^3}\braket{N_B}\,,\label{eq:chiB1}\\[2ex]
\chi^B_2=& \frac{1}{VT^3}\braket{(\delta N_B)^2}\,,\label{eq:chiB2}\\[2ex]
\chi^B_3=&\frac{1}{VT^3}\braket{(\delta N_B)^3}\,,\\[2ex]
\chi^B_4=&\frac{1}{VT^3}\Big(\braket{(\delta N_B)^4}-3\braket{(\delta N_B)^2}^2\Big)\,,\label{eq:chiB4}
\end{align}
with $\braket{\cdots}$ denoting the ensemble average and $\delta N_B=N_B-\braket{N_B}$. Thus the mean value of the net baryon number of the system is given by $M=VT^3\chi_1^{B}$, the variance $\sigma^2=VT^3\chi_2^{B}$, skewness $S=\chi_3^{B}/(\chi_2^{B}\sigma)$, and the kurtosis $\kappa=\chi_4^{B}/(\chi_2^{B}\sigma^2)$, respectively.

%
\begin{figure*}[t]
\includegraphics[width=1\textwidth]{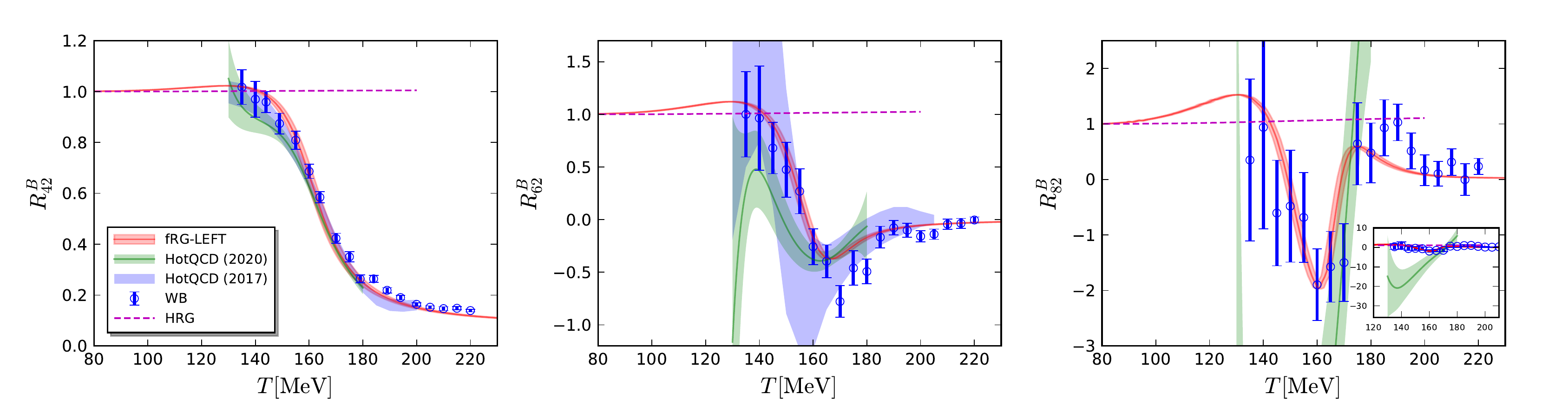}
\caption{$R^{B}_{42}=\chi^{B}_{4}/\chi^{B}_{2}$ (left panel), $R^{B}_{62}=\chi^{B}_{6}/\chi^{B}_{2}$ (middle panel), and $R^{B}_{82}=\chi^{B}_{8}/\chi^{B}_{2}$ (right panel) as functions of the temperature at vanishing baryon chemical potential ($\mu_B=0$). Results from the QCD-assisted LEFT are compared with lattice results from the HotQCD collaboration \cite{Bazavov:2017dus,Bazavov:2017tot,Bazavov:2020bjn} and the Wuppertal-Budapest collaboration (WB) \cite{Borsanyi:2018grb}. The inset in the plot of $R^{B}_{82}$ shows its zoomed-out view. Our results agree quantitatively with the WB-results , and are qualitatively compatible with the HotQCD results. We also compare to the predictions of a hadron resonance gas (HRG) \cite{BraunMunzinger:2003zd}, which predicts only a very mild increase of $R_{n2}^B$ from unity with increasing $T$.
}\label{fig:R42R62R82-T-muB0}
\end{figure*}
%

%
\begin{figure}[b]
\includegraphics[width=0.45\textwidth]{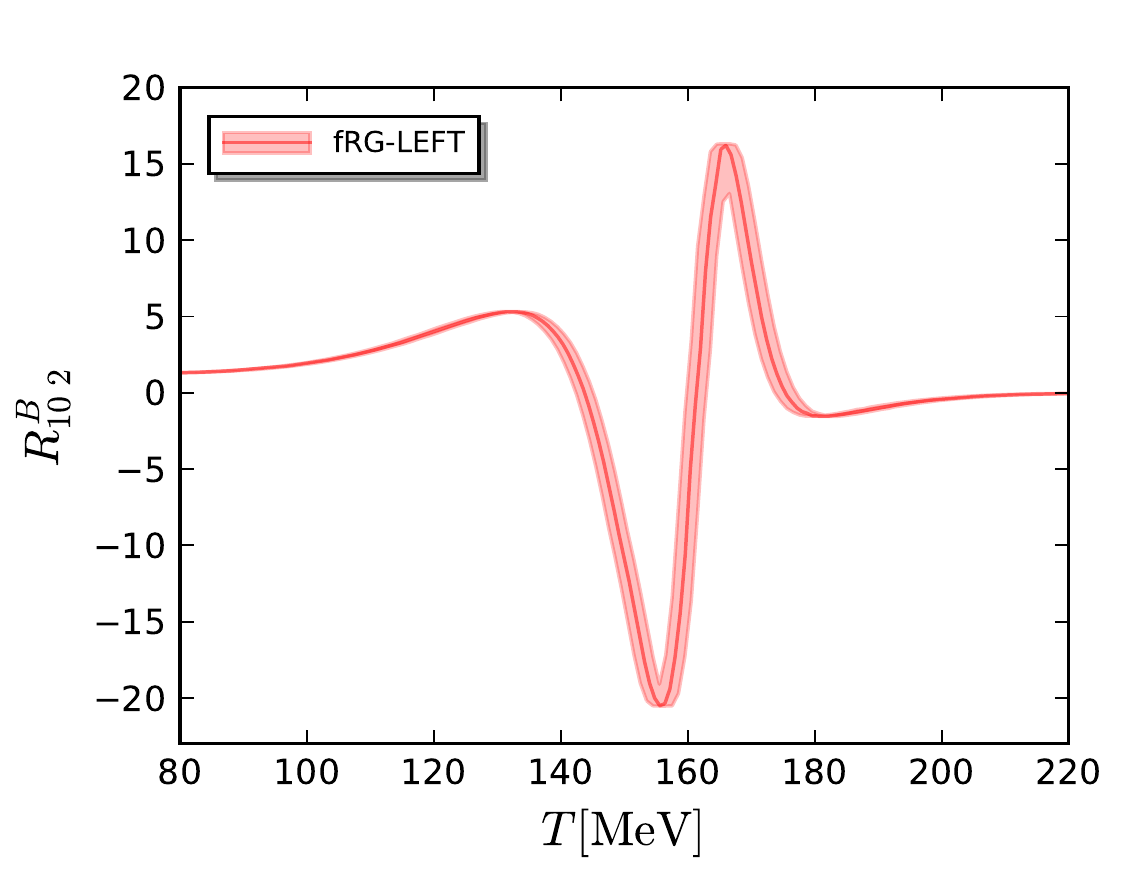}
\caption{$R^{B}_{10\,2}=\chi^{B}_{10}/\chi^{B}_{2}$ as a function of the temperature with $\mu_B=0$ from the QCD-assisted LEFT.
}\label{fig:R102-T-muB0}
\end{figure}
%
In this work the emphasis is, however, put on the baryon number fluctuations of orders higher than the fourth, i.e. $\chi_{n>4}^{B}$, which are named hyper-order baryon number fluctuations. As the low-order ones, the hyper-order susceptibilities are also connected to their respective cumulants, and their relations, taking the fifth through eighth ones for instance, are given as follows, 
{\allowdisplaybreaks
\begin{align}
\chi^B_5=&\frac{1}{VT^3}\Big(\braket{(\delta N_B)^5}-10\braket{(\delta N_B)^2}\braket{(\delta N_B)^3}\Big)\,,\\[2ex]
\chi^B_6=&\frac{1}{VT^3}\Big(\braket{(\delta N_B)^6}-15\braket{(\delta N_B)^4}\braket{(\delta N_B)^2}\nonumber\\[2ex]
&-10\braket{(\delta N_B)^3}^2+30\braket{(\delta N_B)^2}^3\Big)\,,\\[2ex]
\chi^B_7=&\frac{1}{VT^3}\Big(\braket{(\delta N_B)^7}-21\braket{(\delta N_B)^5}\braket{(\delta N_B)^2}\nonumber\\[2ex]
&-35\braket{(\delta N_B)^4}\braket{(\delta N_B)^3}\nonumber\\[2ex]
&+210\braket{(\delta N_B)^3}\braket{(\delta N_B)^2}^2\Big)\,,\\[2ex]
\chi^B_8=&\frac{1}{VT^3}\Big(\braket{(\delta N_B)^8}-28\braket{(\delta N_B)^6}\braket{(\delta N_B)^2}\nonumber\\[2ex]
&-56\braket{(\delta N_B)^5}\braket{(\delta N_B)^3}-35\braket{(\delta N_B)^4}^2\nonumber\\[2ex]
&+420\braket{(\delta N_B)^4}\braket{(\delta N_B)^2}^2\nonumber\\[2ex]
&+560\braket{(\delta N_B)^3}^2\braket{(\delta N_B)^2}-630\braket{(\delta N_B)^2}^4\Big)\,.
\end{align}
}
Different aspects of hyper-order fluctuations have been studied in mean-field approximations in the past, see e.g.\ \cite{Wagner:2009pm, Karsch:2010hm, Schaefer:2011ex}. However, due to the decisive role of non-perturbative quantum fluctuations for these quantities, a treatment beyond mean-field, as in the present work, is necessary for their accurate description. A first study in this direction, discussing hyper-order fluctuations up to $\chi_8$ within a PQM model with the fRG at small $\mu_B/T$ can be found in \cite{Friman:2011pf}.
%
\begin{figure*}[t]
\includegraphics[width=1.\textwidth]{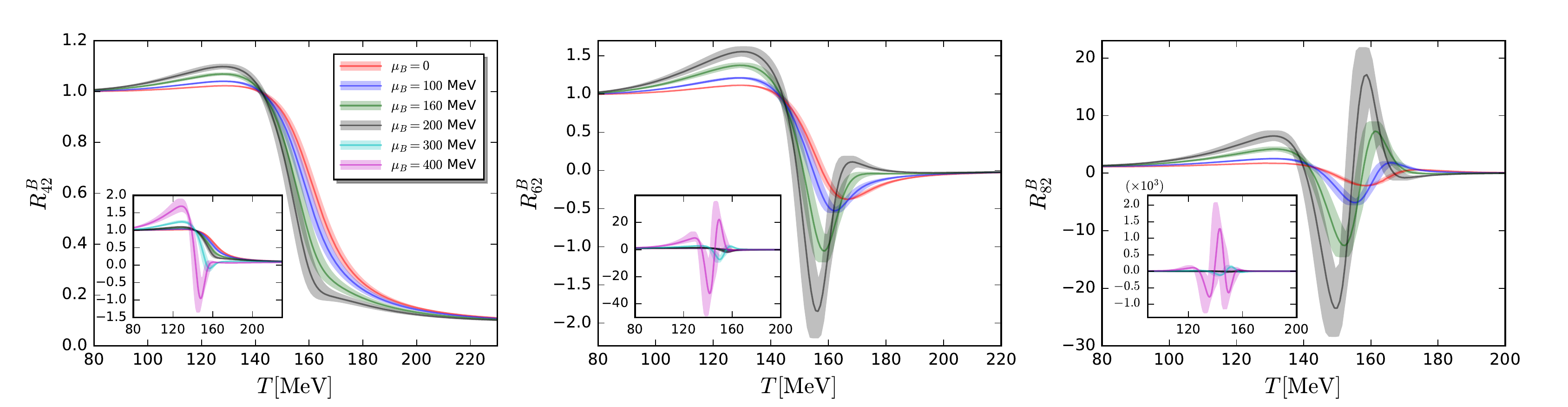}
\caption{$R^{B}_{42}$ (left panel), $R^{B}_{62}$ (middle panel), and $R^{B}_{82}$ (right panel) as functions of the temperature at several values of $\mu_B$. Insets in each plot show their respective zoomed-out view. }\label{fig:R42R62R82-T-muB0to400}
\end{figure*}
%

\section{Numerical results and discussions}
\label{sec:num}

In this section we present and discuss our numerical results for hyper-order fluctuations on the freeze-out curve. At vanishing chemical potential the lower orders are compared to results from lattice calculations. We then discuss the implications of our predictions for the hyper-order baryon number fluctuations for decreasing collision energies (increasing chemical potential) for heavy-ion collision experiments. 


\subsection{Hyper-order baryon number fluctuations at vanishing density: benchmarks and predictions}
\label{subsec:hyper-order0}

We start our discussion of the numerical results in our QCD-assisted low-energy effective theory with benchmark results at vanishing chemical potential, $\mu_B=0$. We have already seen in \sec{subsec:scale} that the fourth order fluctuations $R^B_{42}$, \Eq{eq:Rnm}, agrees quantitatively with the respective lattice result, see \fig{fig:T-adjust} and \fig{fig:R42R62R82-T-muB0}, left panel. We emphasise again  that the thermal dependence of $R^B_{42}$ is a prediction of the present LEFT. Now we also compare the temperature dependence of the hyper-fluctuations $R^B_{62}$ and $R^B_{82}$ with the corresponding lattice results in the middle and right panels of \Fig{fig:R42R62R82-T-muB0}, respectively. We depict both our numerical results and lattice results from the HotQCD collaboration, \cite{Bazavov:2017dus,Bazavov:2017tot,Bazavov:2020bjn}, and the Wuppertal-Budapest collaboration, \cite{Borsanyi:2018grb}. Note that lattice results in \fig{fig:R42R62R82-T-muB0} by the Wuppertal-Budapest collaboration in \cite{Borsanyi:2018grb}, and  $R^B_{62}$ and $R^B_{82}$ by the HotQCD collaboration in \cite{Bazavov:2017dus} are not continuum extrapolated.

With the increase of the order of fluctuations, the uncertainties of lattice results increase significantly. Moreover, the eighth-order fluctuations $R^{B}_{82}$ obtained by the two collaborations show a significant quantitative difference, although their form is qualitatively consistent with each other. 

The hyper-order baryon number fluctuations computed in the current setup are in qualitative agreement with both lattice results. However, our results single out the lattice results of the Wuppertal-Budapest collaboration, with which we observe quantitative agreement. This situation is very reminiscent of the pressure prediction in \cite{Herbst:2013ufa}: similarly to the current situation with lattice predictions of hyper-fluctuations, the pressure predictions of the lattice collaborations had not converged yet. A less advanced version of the current QCD-assisted LEFT framework then predicted the correct pressure result. 
We have also computed the hyper-order fluctuations within a simple hadron resonance gas model \cite{BraunMunzinger:2003zd}. Essentially, they are all constant with only a very minor monotonic increase with $T$ for $T \lesssim 140$~MeV, starting from unity at $T=0$. This is in quantitative agreement with our findings at low temperatures. In summary, the current setup passes all benchmark tests quantitatively and provides the full temperature-dependence of hyper-fluctuations. We have also computed even higher order baryon number fluctuations. In \Fig{fig:R102-T-muB0} we show our result for the temperature-dependence of the tenth order ratio $R^{B}_{10\,2}=\chi^{B}_{10}/\chi^{B}_{2}$ at vanishing chemical potential, $\mu_B=0$. So far no lattice results for the tenth-order fluctuation are available, and the dependence of $R^{B}_{10\,2}$ on the temperature in \Fig{fig:R102-T-muB0} is  a prediction by the current QCD-assisted LEFT and awaits confirmation by other calculations, e.g., lattice QCD, in the future.

%
\begin{figure*}[t]
\includegraphics[width=0.9\textwidth]{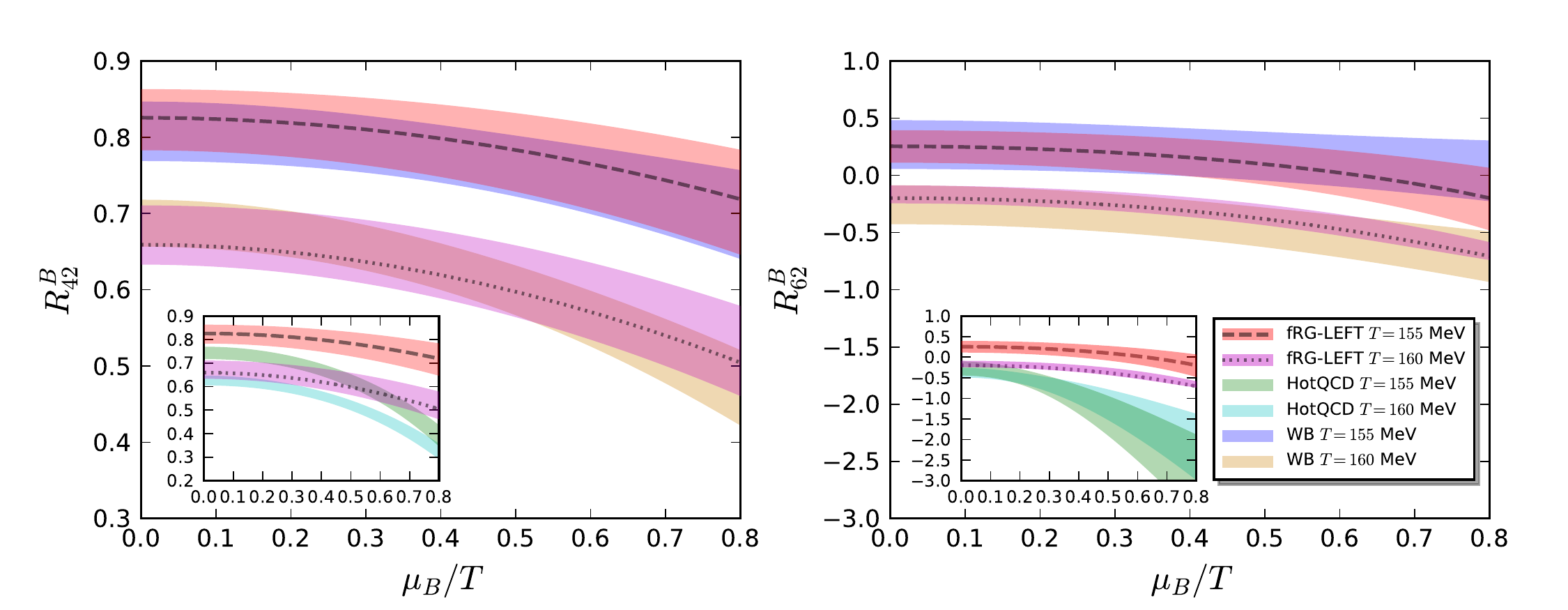}
\caption{$R^{B}_{42}$ (left panel) and $R^{B}_{62}$ (right panel) as functions of $\mu_B/T$ with $T=155$ MeV and $T=160$ MeV from the eighth-order Taylor expansion in $(\mu_B/T)^2$ around vanishing $\mu_B$ shown in \Eq{eq:chiBTay}. Our results from the QCD-assisted LEFT are  compared to those from lattice QCD by the HotQCD collaboration \cite{Bazavov:2020bjn} and the Wuppertal-Budapest collaboration \cite{Borsanyi:2018grb}. We show the comparison to HotQCD in the inlays, as these results deviate considerably from both ours and the WB results.}\label{fig:R42R62-muBoT}
\end{figure*}
%


\subsection{Hyper-order baryon number fluctuations at finite density}
\label{subsec:hyper-ordermuB}
With successfully passing the benchmark tests, we proceed with the results for baryon number fluctuations at finite chemical potential. This will allow us to finally compare the theoretical predictions on the freeze-out curve with the experimental measurements in \sec{subsec:freezout}. 

Equally relevant is the self-consistent evaluation of the reliability of a Taylor expansion in baryon-chemical potential that underlies the extension of lattice results at vanishing chemical potential to $\mu_B\neq 0$. This is particularly important for predictions of the location of the critical end point based on such an expansion. Here we can investigate the reliability range of the Taylor expansion around $\mu_B = 0$ by comparison to our direct computation at finite $\mu_B$.

First we investigate the temperature-dependence of the baryon number fluctuations for different chemical potentials. This also allows us to discuss the reliability bounds of the current LEFT-setup for increasing chemical potential. In \Fig{fig:R42R62R82-T-muB0to400} we show the temperature-dependence of the ratios $R^{B}_{42}$, $R^{B}_{62}$ and $R^{B}_{82}$ for chemical potentials $\mu_B=0, 100, 160, 200, 300, 400$ MeV. First, we note that at small temperatures the thermodynamic properties of the QCD medium are well described by a dilute gas of hadrons, where the net-baryon number follows a Skellam distribution. Thus, all ratios approach unity at vanishing temperature. At very large temperature the system is governed by asymptotically free quarks, where fluctuations approach to the trivial Stefan-Boltzmann limit, and $R^{B}_{n2}$ goes to zero for all $n>4$ at large $T$. Consequently, the non-trivial behaviour of the fluctuations between these two limiting cases shown in \Fig{fig:R42R62R82-T-muB0to400} is directly related to the crossover from the hadronic- to the quark-gluon regime of QCD. The magnitude, but also the error of the fluctuations grow with increasing chemical potential. Both effects are more pronounced for higher order fluctuations. The increase in magnitude is directly linked to the sharpening of the chiral crossover with increasing chemical potential, cf.\ \fig{fig:QCD-scalematching}. We expect that the current LEFT-setup is gradually loosing its predictive power for fluctuations on the freeze-out curve due to the rapid increase of the computational error for higher-order fluctuations at large baryon chemical potential, e.g., $R^{B}_{82}$ with $\mu_B\gtrsim 200$ MeV. All results of the subsequent investigations have to be evaluated with this estimate on our systematic error.  
	
%
\begin{figure*}[t]
\includegraphics[width=0.9\textwidth]{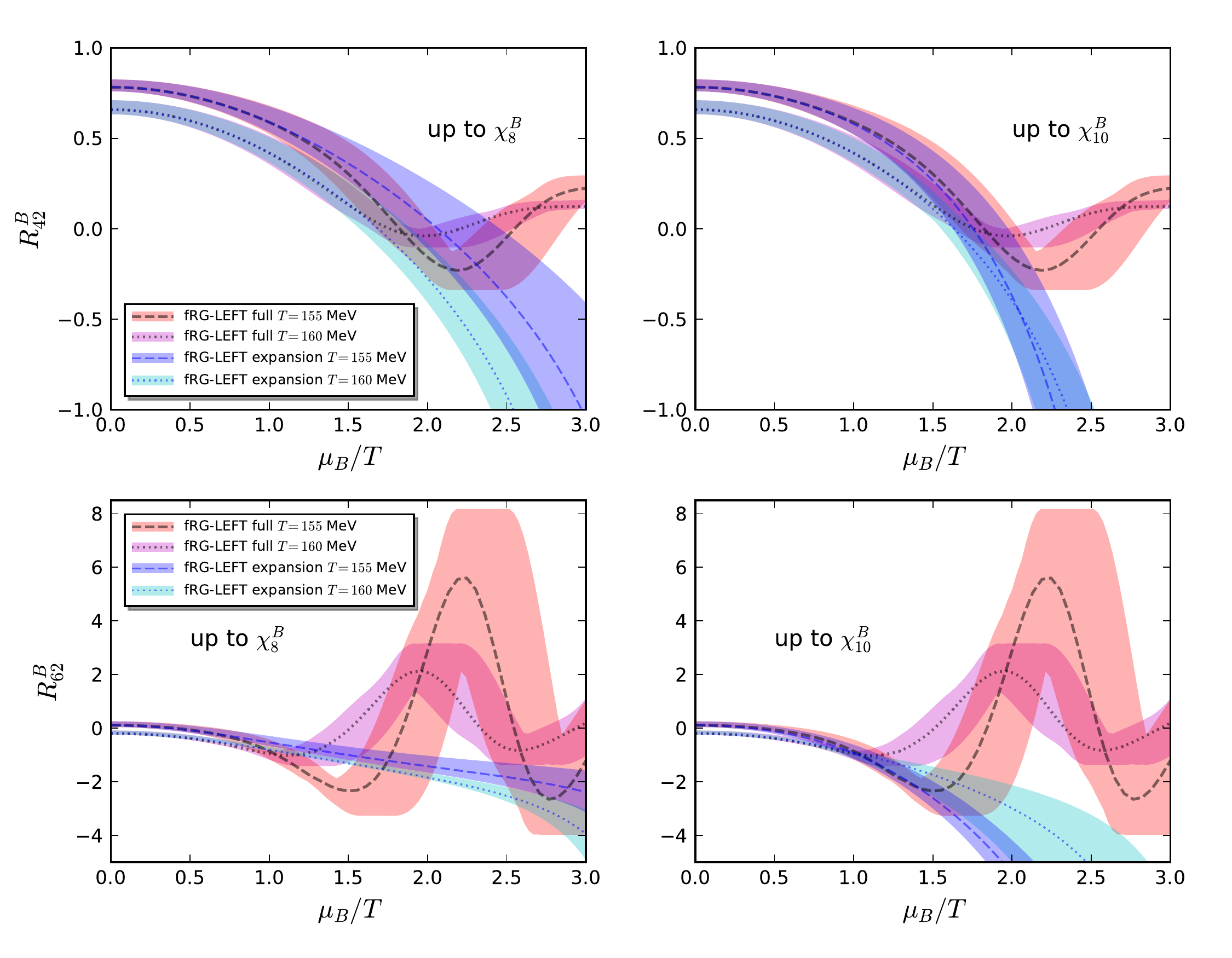}
\caption{Comparison between the direct calculation of baryon number fluctuations $R^{B}_{42}$ (upper panels) and $R^{B}_{62}$ (lower panels) via \eq{eq:suscept} at finite $\mu_B$ and the Taylor expansion up to $\chi^B_8(0)$ in \eq{eq:chiBTay} (left panels) and to $\chi^B_{10}(0)$ (right panels). Both calculations are performed within the QCD-assisted LEFT used in the present work. $R^{B}_{42}$, $R^{B}_{62}$ are shown as functions of $\mu_B/T$ with $T=155$ MeV and $T=160$ MeV.}\label{fig:R42R62expansion-muBoT}
\end{figure*}
%

For the evaluation of the reliability regime of the Taylor expansion about vanishing chemical potential we consider the Taylor expansion of the pressure in \Eq{eq:pres} in powers of $\hat{\mu}_{B}\equiv\mu_B/T$ around $\hat{\mu}_{B}=0$. This leads us to 
\begin{align}
\frac{p(\mu_B)}{T^4}&=\frac{p(0)}{T^4}+\sum_{i=1}^{\infty}\frac{\chi^B_{2i}(0)}{(2i)!}\,\hat{\mu}_{B}^{2i}\,,\label{eq:cmu}
\end{align}
with the expansion coefficients $\chi^B_{2i}(0)=\chi^B_{2i}(\mu_B=0)$, the hyper-order fluctuations of the baryon charge. In \Eq{eq:cmu} we have suppressed the temperature-dependence of all functions for the sake of readability. Truncating the Taylor expansion in \Eq{eq:cmu} at the eighth order, $\hat{\mu}_{B}^{8}$, and employing \eq{eq:suscept}, we obtain the expanded baryon number fluctuations, 
\begin{align}
\chi^B_2(\mu_B)\simeq&\chi^B_2(0)+\frac{\chi^B_4(0)}{2!}\hat{\mu}_{B}^{2}+\frac{\chi^B_6(0)}{4!}\hat{\mu}_{B}^{4}+\frac{\chi^B_8(0)}{6!}\hat{\mu}_{B}^{6}\,,\nonumber \\[2ex]
\chi^B_4(\mu_B)\simeq&\chi^B_4(0)+\frac{\chi^B_6(0)}{2!}\hat{\mu}_{B}^{2}+\frac{\chi^B_8(0)}{4!}\hat{\mu}_{B}^{4}\,,\nonumber\\[2ex]
\chi^B_6(\mu_B)\simeq&\chi^B_6(0)+\frac{\chi^B_8(0)}{2!}\hat{\mu}_{B}^{2}\,.\label{eq:chiBTay}
\end{align}
In \Fig{fig:R42R62-muBoT} we show the ratios $R^B_{42}=\chi^B_4/\chi^B_2$ and $R^B_{62}=\chi^B_6/\chi^B_2$ based on the Taylor expansion for two fixed temperatures:  $T=155$\,MeV (close to the crossover temperature $T_c$ at $\mu_B=0$) and $T=160$\,MeV (slightly above $T_c$). As an input we use $\chi^B_{2i}(0)$ ($i \!=\! 1,\, 2,\, 3,\, 4$) from the current setup as well as from the lattice (HotQCD collaboration \cite{Bazavov:2020bjn} and Wuppertal-Budapest collaboration \cite{Borsanyi:2018grb}), depicted already in \Fig{fig:R42R62R82-T-muB0}. As expected, the LEFT-results for the $\mu_B$-dependence of $R^{B}_{42}$ and $R^{B}_{62}$ agrees qualitatively with both lattice results. Moreover, it agrees quantitatively with the Wuppertal-Budapest result. Note that constraints, e.g., strangeness neutrality or a fixed ratio of the electric charge to the baryon number density, are not implemented in all the results in \Fig{fig:R42R62-muBoT}. For more details about effects of these constraints on the fluctuations and correlations of conserved charges, see the relevant discussions in, e.g., \cite{Bazavov:2017dus,Bazavov:2020bjn} in lattice QCD and \cite{Fu:2018qsk,Fu:2018swz} in fRG.

Since we are not restricted by a sign problem within the fRG approach, the $\chi^B_n(\mu_B)$'s in \Eq{eq:suscept} can also be computed directly for the current QCD-assisted LEFT without resorting to a Taylor expansion. By comparing this to the results of the Taylor expansion, we can study its range of validity. The results are presented in the left panel of \Fig{fig:R42R62expansion-muBoT}, again for $T=155$\,MeV and $T=160$\,MeV and with the Taylor expansion up to eights order in $\hat \mu_B$.

We observe that the result for $R^{B}_{42}$ from the Taylor expansion in \Eq{eq:chiBTay} agrees quantitatively with that from the full calculation for $\mu_B/T\lesssim 1.2$ for $T=155$\,MeV and $\mu_B/T\lesssim 1.5$ for $T=160$\,MeV. Not surprisingly, this reliability regime is reduced significantly for the hyper-order fluctuation $R^{B}_{62}$ to $\mu_B/T\lesssim 1.2$ for $T=160$\,MeV and $\mu_B/T\lesssim 0.8$ for $T=155$\,MeV. 
For larger $\mu_B/T$, the fluctuations show a non-trivial oscillatory behaviour that cannot be captured by a (low-order) Taylor expansion. We emphasise that this is not an artefact of our model, but rather a generic, physical feature of these fluctuation observables. It reflects the increasingly pronounced non-monotonic temperature dependence of $R^{B}_{n2}$ due to long-range correlations in the crossover region, as seen in \Fig{fig:R42R62R82-T-muB0to400}. In particular, $R^{B}_{n2}$ develops distinctive areas around the crossover at larger $\mu_B$ where its value varies significantly, even including sign changes. By following a line of fixed $T$ close to $T_c$ and increasing $\mu_B$ in the phase diagram, these areas are crossed eventually, leading to the oscillatory behaviour seen in \Fig{fig:R42R62expansion-muBoT}. This is also evident in the right plot of \Fig{fig:phasediagram}, where we show the magnitude of $R^{B}_{42}$ in the phase diagram. Since these strong fluctuations only occur at larger $\mu_B$, the resulting characteristic qualitative features cannot be captured by a (low-order) Taylor expansion at $\mu_B=0$; it is bound to fail at the onset of the oscillatory behaviour, i.e.\ around $\mu_B/T \gtrsim 1$. 

It is also interesting to evaluate to what extend a higher-order Taylor expansion can improve its reliability. We therefore include our prediction for $R^B_{10\, 2}$ from \fig{fig:R102-T-muB0}, hence extending the Taylor expansion in \Eq{eq:chiBTay} to the tenth order. The resulting comparison is shown in right panel of \Fig{fig:R42R62expansion-muBoT}. Interestingly, this does not change the compatibility regime for $T=160$\,MeV significantly. In turn, there are significant changes for $T=155$\,MeV. While the deviations for $R_{42}^B$ 
start to grow at roughly the same $\mu_B/T$ as for the eights-order expansion, that is $\mu_B/T\lesssim 1.2$, the result for $R_{42}^B$ stays compatible with the full result for larger values. For $R_{62}^B$ the reliability regime is essentially doubled: it rises from $\mu_B/T\lesssim 0.8$ to $\mu_B/T\lesssim 1.5$. 

The analysis above suggests the following picture: We have a temperature-dependent reliability range of the Taylor expansion, 
\begin{align}\nonumber 
 T=155\,\textrm{MeV}:\quad &[\mu_B/T]_\textrm{Max}\approx 1.5\,,\\[1ex] T=160\,\textrm{MeV}:\quad &[\mu_B/T]_\textrm{Max}\approx 1.2\,.
\label{eq:RadiusConverge}\end{align}
Moreover, the results have already converged for $R_{42}^B,R_{62}^B$ for $T=160$\,MeV as well as for $R^B_{42}$ for $T=155$\,MeV within the eighth order and for $\mu_B/T\lesssim [\mu_B/T]_\textrm{Max}(T)$. In turn, convergence for $R^B_{62}$ for $\mu_B/T\lesssim  [\mu_B/T]_\textrm{Max}$ requires the tenth order Taylor expansion for $T=155$\,MeV. Note that the values of $[\mu_B/T]_\textrm{Max}$ in \Eq{eq:RadiusConverge} might be increased a bit when terms of orders higher than the tenth one are included in the Taylor expansion. However, as we have discussed above, the full results of the baryon number fluctuations show a non-trivial oscillatory behaviour, which is generic, and stems from the fact that the system crosses over different phases with the increase of $\mu_B$, cf. the right panel of \Fig{fig:phasediagram}. 
For a discussion of the radius of convergence based on mean-field theory we refer to \cite{Karsch:2010hm}. In general, it is given by the distance to the nearest singularity of the equation of state in the complex $\mu_B/T$ plane. Hence, possible candidates for this singularity are the CEP, the Roberge--Weiss endpoint at imaginary $\mu_B$ \cite{Roberge:1986mm}, or the Yang--Lee edge singularity in the complex plane \cite{Yang:1952be}. Evidently, the distance to the CEP is far larger than the radius estimated here. In turn, the closest endpoint at imaginary chemical potential is at $|\mu_B/T|\leq \pi$. For physical quark masses it is most probably close to $|\mu_B/T|=\pi$. For a discussion within QCD-flows see~\cite{Braun:2009gm}, for lattice results see e.g.~\cite{Philipsen:2014rpa}. Hence, a particularly intriguing option is the Yang--Lee edge singularity, for a discussion see e.g.~\cite{Stephanov:2006dn, Mukherjee:2019eou}. However, while the location of the edge singularity has been determined for critical $O(N)$ theories \cite{Connelly:2020gwa}, it is still unknown for QCD.

This interpretation also implies that the results from the Taylor expansion fail to agree even qualitatively with the correct $\mu_B/T$-dependence for $\mu_B/T_c \gtrsim [\mu_B/T]_\textrm{Max}(T)$, see  \fig{fig:R42R62expansion-muBoT}. Interestingly, $[\mu_B/T]_\textrm{Max}(T)$ seems to grow for smaller temperatures. Whether or not this holds true requires a more systematic study, which will be considered elsewhere. In conclusion, the extrapolation of fluctuations of conserved charges in the vicinity of the chiral crossover within a Taylor expansion loses its predictive power for $\mu_B\gtrsim 200$\,MeV, at least at tenth order.  

%
\begin{figure*}[t]
\includegraphics[width=1.\textwidth]{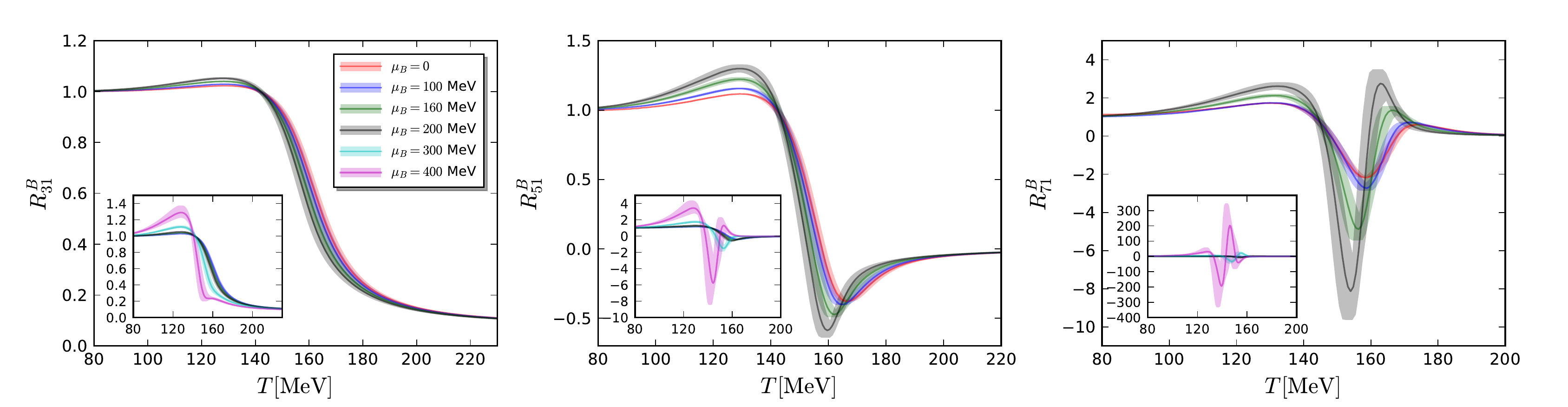}
\caption{$R^{B}_{31}$ (left panel), $R^{B}_{51}$ (middle panel), and $R^{B}_{71}$ (right panel) as functions of the temperature at several values of $\mu_B$. Insets in each plot show their respective zoomed-out view. }\label{fig:R31R51R71-T-muB0to400}
\end{figure*}
%

\begin{figure}[b]
\includegraphics[width=0.45\textwidth]{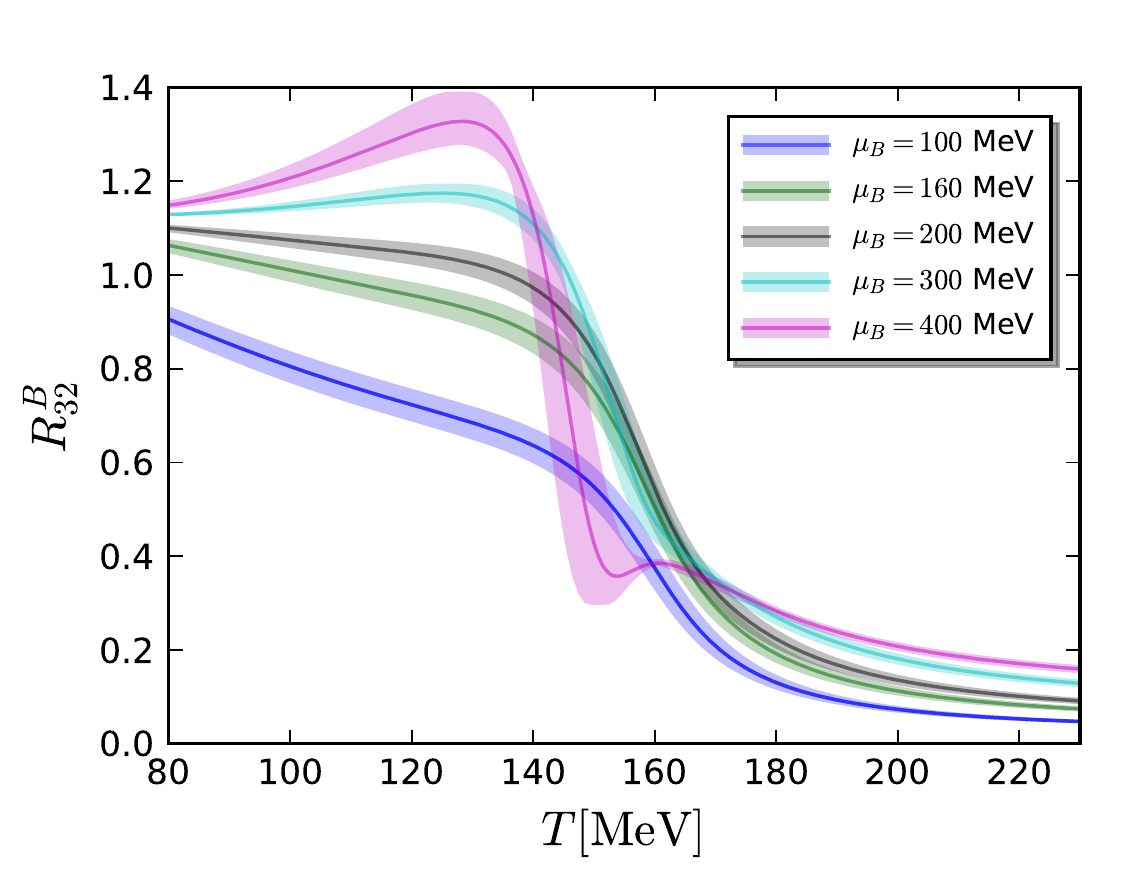}
\caption{Baryon number fluctuation $R^{B}_{32}$ as a function of the temperature at several values of $\mu_B$. }\label{fig:R32-T-muB0to400}
\end{figure}

In \fig{fig:R31R51R71-T-muB0to400} we show our full results for the temperature-dependence of $R^B_{31}$, $R^B_{51}$, and $R^B_{71}$ with different values of baryon chemical potential. A further relevant odd fluctuation observable is $R^B_{32}$, depicted in \fig{fig:R32-T-muB0to400}. Its experimental analogue, the proton number fluctuation $R^p_{32}$ has been already measured in Au+Au central (0-5\%) collisions at STAR, a comparison will be presented and discussed in \sec{sec:CEP}.


%
%


\subsection{Determination of the  freeze-out curve}
\label{subsec:freezeoutCurve}	

%
\begin{figure*}[t]
\includegraphics[width=0.42\textwidth]{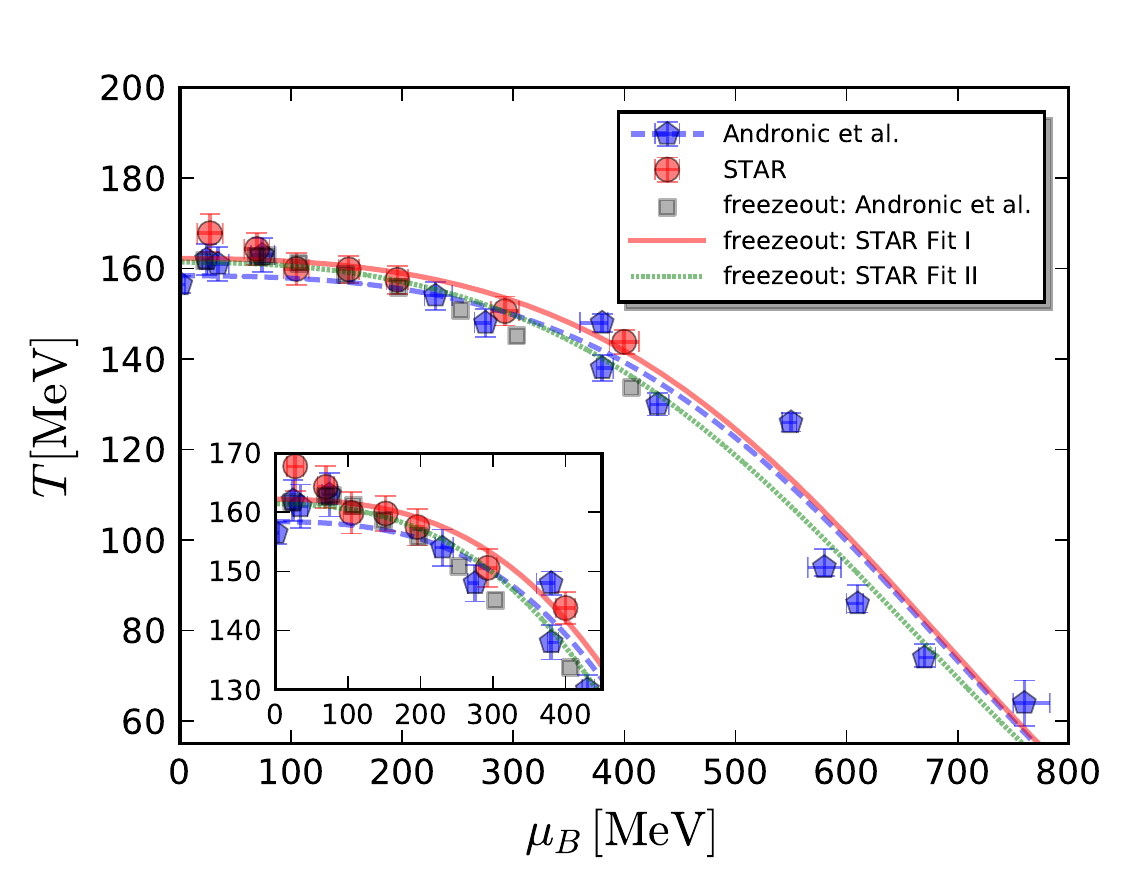}
\includegraphics[width=0.55\textwidth]{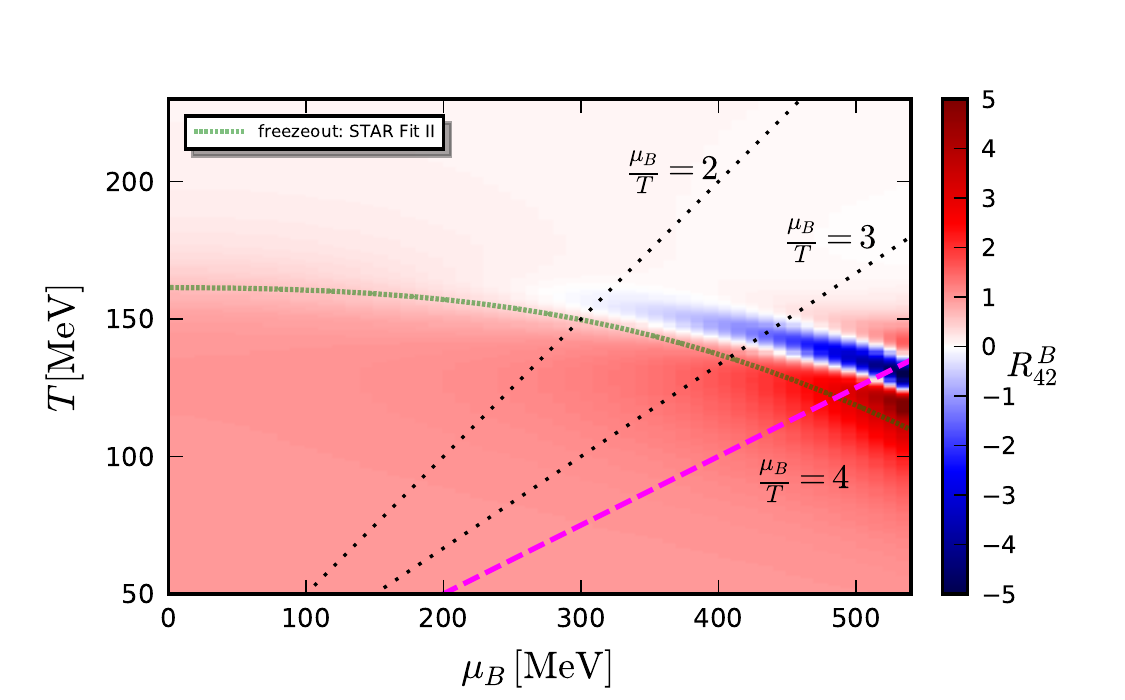}
\caption{Left panel: chemical freeze-out temperature and baryon chemical potential in the $T\!-\!\mu_B$ plane. The blue pentagons and red circles show the freeze-out data from Andronic {\it et al.} \cite{Andronic:2017pug} and STAR experiment \cite{Adamczyk:2017iwn}, respectively. The blue dashed line represents the parametrisation of blue pentagons through \eq{eq:muBCFparatri} and \eq{eq:TCFparatri}. The red solid and green dotted lines show the parametrisation of the STAR data based on all the seven data points, and only the four data points in the middle region ($100\,\mathrm{MeV}\lesssim\mu_B\lesssim 300\,\mathrm{MeV}$), respectively. The grey squares are obtained by interpolating the blue pentagons. The inlay zooms in the low-$\mu_B$ region. \\ 
Right panel: Baryon number fluctuations $R^{B}_{42}$ in the $T\!-\!\mu_B$ plane. The freeze-out curve is the STAR Fit II. The dashed line at $\mu_B/T=4$ constitutes the reliability bound of the computations in \cite{Fu:2019hdw} based on the potential emergence of new degrees of freedom discussed in \cite{Fu:2019hdw, Braun:2019aow, Fischer:2018sdj}. The dashed lines at $\mu_B/T=2,3$ are reliability estimates of lattice results as well as old ones from functional approaches, see also \Fig{fig:QCD-scalematching}.}\label{fig:phasediagram}
\end{figure*}
%

The quantitatively successful benchmark tests analysed in \sec{subsec:hyper-order0}, and the evaluation of baryon number fluctuations at finite chemical potential in \sec{subsec:hyper-ordermuB} allow us to discuss our main goal: the comparison of theoretical predictions on the baryon number fluctuations with experimental measurements. 

A direct comparison between theory and experiment is a very  challenging task. This is due to the fact that experimental data are affected by many factors. First, this concerns the acceptance of the detector such as the transverse momentum $p_T$ range, rapidity window and the centrality dependence, e.g.\ \cite{Adamczyk:2013dal,Luo:2015ewa,Bzdak:2016sxg, He:2017zpg,Adam:2020unf,STAR:2021rls}, see \cite{Luo:2017faz,Adamczyk:2017iwn} for more details. Second, the physics setup used in theory and experiment may differ by the presence of volume fluctuations, e.g.\  \cite{Luo:2013bmi,Chatterjee:2019fey,Chatterjee:2020nnn}, finite volume effects on the location of the chiral phase boundary, e.g.\ \cite{Braun:2010vd, Braun:2011iz, Tripolt:2013zfa, Almasi:2016zqf, Klein:2017shl, Li:2017zny, Liu:2020elq, Wan:2020vaj}, the question of global baryon number conservation, e.g.\  \cite{He:2016uei,Braun-Munzinger:2016yjz,Vovchenko:2020tsr}, the inclusion of resonance decays, e.g.\  \cite{Nahrgang:2014fza,Zhang:2019lqz}, and others. 

All these different effects and experimental restrictions give rise to  non-critical contributions to fluctuation observables in experiments, and pinning down their contributions plays a pivotal role in identifying the critical signals in the BES experiment. Additionally, due to critical slowing down, non-equilibrium effects become important in the vicinity of the CEP \cite{Berdnikov:1999ph}, which necessitates a theoretical description of the dynamics of critical fluctuations. For more details about recent progress on the dynamics of critical fluctuations in QCD, see \cite{Bluhm:2020mpc} and references therein. We emphasise, however, that the present results of QCD-assisted LEFT model are well outside the critical region. Therefore they are not subject to critical scaling in the vicinity of the CEP. 

In this work we will not take into account the non-critical and dynamical effects discussed above. Instead, we assume that the measured cumulants of the net-proton multiplicity distribution at a given collision energy are in one-to-one correspondence to the calculated fluctuations in \Eq{eq:suscept} with single values for $T$ and $\mu_B$ (with other collision parameters e.g., the centrality and rapidity range fixed). Then, it is suggestive to identify the values of $T$ and $\mu_B$ with the ones when the chemical freeze-out occurs, viz. $T_{_{\textrm{CF}}}$ and ${\mu_B}_{_{\textrm{CF}}}$. Such an approach for the comparison is usually employed in fluctuation studies of equilibrium QCD matter within functional methods or lattice simulations, see e.g.\ \cite{Fu:2015amv, Fu:2016tey, Almasi:2017bhq, Isserstedt:2019pgx, Bazavov:2020bjn}.

We adopt the freeze-out temperatures and baryon chemical potentials from \cite{Andronic:2017pug} and from the STAR experiment \cite{Adamczyk:2017iwn}, which are shown in the left panel of \Fig{fig:phasediagram} by the blue pentagons and red circles, respectively. They are both obtained from the analysis of hadron yields in the statistical hadron resonance gas model, see the  aforementioned references for more details. The freeze-out data in \cite{Andronic:2017pug} has also been parametrised as functions of the collision energy as follows 
\begin{subequations}\label{eq:freeze-outFit}
\begin{align}
{\mu_B}_{_{\textrm{CF}}}&=\frac{a}{1+0.288\sqrt{s_{\mathrm{NN}}}},\label{eq:muBCFparatri}
\end{align}
with $a=1307.5$ MeV, and
\begin{align}
T_{_{\textrm{CF}}}&=\frac{T^{(0)}_{_{\textrm{CF}}}}{1+\exp\big(2.60-\ln(\sqrt{s_{\mathrm{NN}}})/0.45\big)},\label{eq:TCFparatri}
\end{align}
\end{subequations}
with $T^{(0)}_{_{\textrm{CF}}}=158.4$ MeV. This parametrisation is depicted with the blue dashed line in the left panel of \Fig{fig:phasediagram}. We use the same parametrisation functions in \Eq{eq:freeze-outFit} to fit the freeze-out data in STAR experiment, i.e., the red circle points. For this fit we invoke two procedures, called STAR Fit I \& II in the following: \\[-2ex]

For the first one, STAR Fit I, we simply take all 7 data points. The corresponding freeze-out curve is depicted by the red solid line in the left panel of \fig{fig:phasediagram}. 

%
\begin{figure*}[t]
\includegraphics[width=0.9\textwidth]{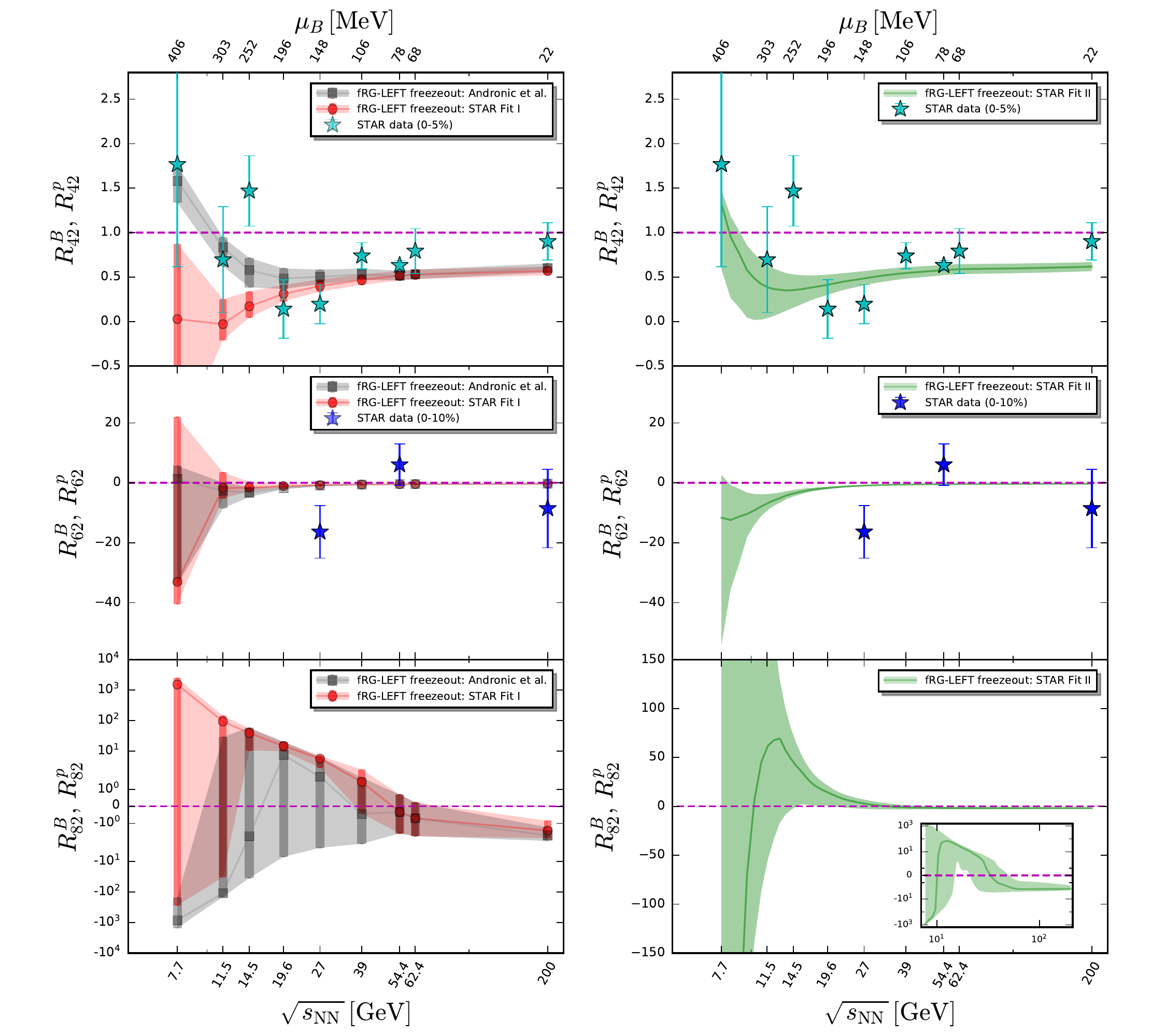}
\caption{QCD-assisted LEFT (fRG-LEFT): Baryon number fluctuations $R^{B}_{42}$ (top), $R^{B}_{62}$ (middle), and $R^{B}_{82}$ (bottom) as functions of the collision energy. 
Left panels: the freeze-out points are those from Andronic {\it et al.} \cite{Andronic:2017pug} (grey) and the STAR experiment \cite{Adamczyk:2017iwn} (red).
Right panels: The freeze-out curve, STAR Fit II, is obtained from the freeze-out parameters of the STAR experiment \cite{Adamczyk:2017iwn}. 
The theoretical error bands show a highly correlated error, and should be interpreted as a family of curves with the same qualitative behaviour as the central curve. For more explanations see \sec{subsec:freezeoutCurve} with \fig{fig:phasediagram}. 
\newline
STAR data: $R^{p}_{42}$ (top) is the kurtosis of the net-proton distributions measured in Au+Au central (0-5\%) collisions \cite{Adam:2020unf}. $R^{p}_{62}$ (middle) is the result on the six-order cumulant of the net-proton distribution at $\sqrt{s_{\mathrm{NN}}}$=200 GeV, 54.4 GeV and 27 GeV with centrality 0-10\% \cite{STAR:2021rls}. 
}\label{fig:Rm2-sqrtS}\vspace{-0.5cm}
\end{figure*}
%
For the construction of the second one, STAR Fit II, we shall argue that some of the data points are potentially flawed, or rather await a physics explanation, and should be dropped accordingly in a fit based on \eq{eq:freeze-outFit}. Accordingly, we drop the first two data points at small chemical potential as well as the last one at the largest available chemical potential $\mu_B\sim 400$. From general considerations we do not expect the freeze-out curve to rise with increasing chemical potential. Moreover, the physically motivated fit formula does not describe sign-changes of the curvature of the freeze-out curve. For a respective discussion and possible explanation for the only apparent rise see \cite{Bluhm:2020rha}. The last data point is also not well-described by the fitting procedure described here. This may indicate the onset of a regime with different physics/phases. In this case, \Eq{eq:freeze-outFit} would not be an appropriate fit function. It may also indicate the onset of a regime of rapidly worsening systematics. In this case more points are needed in this regime.  

The freeze-out line of STAR Fit II is depicted by the green dotted line in the left panel of \fig{fig:phasediagram}. In comparison to STAR Fit I, STAR Fit II is located at slightly lower temperatures, which is more pronounced when $\mu_B\gtrsim 200$\,MeV. In the right panel of \Fig{fig:phasediagram}, we show the baryon number fluctuation $R^{B}_{42}$ in the $T\!-\!\mu_B$ plane. It can be observed that a narrow blue band, indicating the regime of negative $R^{B}_{42}$, develops around the crossover starting at $\mu_B\sim 250$ MeV. The freeze-out curve STAR Fit II is approaching towards the boundary of the blue region firstly at small $\mu_B$, and then deviates a bit from it at large $\mu_B$. We emphasise that the large chemical potential region, and in particular  asymptotically large $\mu_B\gtrsim 500$ MeV, is beyond of the reliability bound of the current computation, $\mu_B/T=4$, see \Fig{fig:QCD-scalematching}. For a detailed discussion see \sec{subsec:scale}.


\subsection{Hyper-order baryon number fluctuations on the  freeze-out curve}
\label{subsec:freezout}	

The determination of the freeze-out curve completes our setup, which enables us to compute hyper-order baryon number fluctuations $R_{nm}^B$ along the freeze-out line within the QCD-assisted LEFT. These results are then used to compare with the experimental measurements of cumulants $R_{nm}^p$ of the net-proton distribution from STAR experiment. 

Before we discuss the numerical results, we also emphasise once more, that it follows from the analysis of \sec{subsec:hyper-ordermuB}, that the simple extrapolation with the Taylor expansion about $\mu_B=0$ lacks predictive power for $\mu_B\gtrsim 250$\,MeV, that is $\sqrt{s_{\rm NN}}\lesssim 15$\,GeV, see \Eq{eq:RadiusConverge}. Moreover, it even lacks predictive power for the qualitative behaviour. 

In the left panel of \Fig{fig:Rm2-sqrtS} we show the $\sqrt{s_{\rm NN}}$- or chemical potential dependence of the baryon number fluctuations $R^{B}_{42}$, $R^{B}_{62}$, and $R^{B}_{82}$ for the freeze-out lines from Andronic {\it et al.} \cite{Andronic:2017pug} and STAR Fit I. The freeze-out line from Andronic {\it et al.} is obtained from an interpolation of the freeze-out data, the grey squares in \fig{fig:phasediagram}. 

In the right panel of \Fig{fig:Rm2-sqrtS} we show the same observables for the freeze-out line of STAR Fit II. As discussed in \sec{subsec:freezeoutCurve}, we have singled out the results for this freeze-out curve as the best-informed computation. 

In both panels of \fig{fig:Rm2-sqrtS} we also show the experimental measurement of cumulants of the net-proton distributions in the beam energy scan experiments from the STAR collaboration. The fourth-order fluctuations, $R^{p}_{42}$, of the net-proton multiplicity distributions are measured in Au+Au collisions with centrality 0-5\%, transverse momentum range $0.4< p_T\,(\mathrm{GeV}/c)\,<2.0$, and rapidity $|y|<0.5$, cf. \cite{Adam:2020unf} for more details. Moreover, results for the sixth-order cumulant of the net-proton distribution, $R^{p}_{62}$, are also presented in the middle plot of \Fig{fig:Rm2-sqrtS}, which are obtained at three values of the collision energy, i.e., $\sqrt{s_{NN}}$=200 GeV, 54.4 GeV and 27 GeV with centrality 0-10\% \cite{STAR:2021rls}. 

The theoretical results for the fourth-order fluctuations $R^B_{42}$ from the present QCD-assisted LEFT for all freeze-out curves considered are compatible with the respective experimental measurement of the $\kappa\sigma^{2}$ of net-proton distributions in 0-5\% central Au+Au collisions. 
In particular, the theoretical  results feature a  non-monotonic $\sqrt{s_{\rm NN}}$-dependence: $R^B_{42}$ first decreases with decreasing beam energy and then increases. The details of this behaviour, in particular how pronounced it is, is highly sensitive to the precise location of the freeze-out. For example, the increase at small $\sqrt{s_{\rm NN}}$ is larger for smaller freeze-out temperatures. Thus, the weak increase for STAR Fit 1 originates in the slightly larger freeze-out temperature of this freeze-out fit. 
This shows that even small variations in the freeze-out temperature have a substantial effect on the fluctuations in this regime. The underlying reason is that the freeze-out happens in or close to the crossover region, where the fluctuations vary significantly, see \Fig{fig:R42R62R82-T-muB0to400}. Importantly, this regime cannot be accessed within the extrapolation of the Taylor expansion at least within the current order.

This entails that extrapolations based on a Taylor expansion are bound to fail to describe the data in this regime reliably. Consequently this calls for qualitatively improved direct theoretical computations at small beam-energies. This is work in progress and we hope to report on the respective results soon. 

Our results for the hyper-order fluctuations $R^B_{62}$ and $R^B_{82}$ are shown in the middle and bottom panel of \Fig{fig:Rm2-sqrtS}. For small chemical potentials or large collision energies both fluctuation observables are negative. Moreover, $R^B_{62}$ decreases with decreasing $\sqrt{s_\mathrm{NN}}$, while $R^B_{82}$ increases. The occurrence of non-monotonicities of $R^B_{62}$ and $R^B_{82}$ at lower beam energies cannot be shown within the accuracy limits of the current study.

For $\sqrt{s_\mathrm{NN}}$=200 GeV, 54.4 GeV and 27 GeV we can compare our results for $R^{B}_{62}$ to STAR data within 0-10\% centrality \cite{STAR:2021rls}. One observes that our results are in agreement with the experimental data within errors at $\sqrt{s_\mathrm{NN}}$=200 GeV and 54.4 GeV, though the central value of STAR data at $\sqrt{s_\mathrm{NN}}$=54.4 GeV is positive. Both the theory and experiment show negative values for the sixth-order fluctuations at the collision energy $\sqrt{s_\mathrm{NN}}$=27 GeV.

Another interesting property of the current LEFT setting is that the non-monotonic behaviour of our results for $R^{B}_{n2}$ at large $\mu_{B}$ in \Fig{fig:Rm2-sqrtS} does not arise from critical physics: in the LEFT used here, the CEP is at significantly larger $\mu_B\gtrsim 700$\,MeV. Moreover, it is well established that the critical region is only very small. It is already small within mean-field approximations of low-energy effective theories, and additionally shrinks considerably if quantum, thermal and density equilibrium fluctuations are taken into account, see \cite{Schaefer:2006ds}. Moreover, this does not change if transport processes are taken into account, see \cite{Bluhm:2018qkf}. 

In the present LEFT the increasing trend at large $\mu_{B}$ region originates from two effects: First, fluctuations are enhanced since the chiral crossover becomes sharper with increasing $\mu_B$. This leads to a stronger non-monotonic behaviour of $R^{B}_{42}$ as a function of $T$, see \Fig{fig:R42R62R82-T-muB0to400}. This sharpening is also present in the vicinity of a CEP. Second, the freeze-out temperature is shifted away from the pseudo-critical temperature towards small beam-energies, thereby probing different regimes of the cumulants. However, it should be noted that the uncertainty of our results increases significantly in the low energy region. These uncertainties include an estimate for the systematic error of the QCD-assisted LEFT-approach. This systematic error stems from the uncertainty in the matching of the in-medium scales of the LEFT and QCD, encoded in the coefficients $c_{_{T}}$ and $c_{\mu}$ in \Eq{eq:rescale}. Moreover, for larger chemical potential the current LEFT lacks the back-reaction of the $\mu_B$-dependence of the glue-dynamics. While inherently small, it might still play a r$\hat{\textrm{o}}$le. Furthermore, it is found that $R^{B}_{42}$ can also be suppressed in the regime of low collision energies due to the effect of global baryon number conservation, cf. \cite{Braun-Munzinger:2016yjz,Braun-Munzinger:2020jbk,Vovchenko:2020tsr} which will be included in our future work.


\subsection{Search for the CEP} \label{sec:CEP}

%
\begin{figure*}[t]
\includegraphics[width=0.48\textwidth]{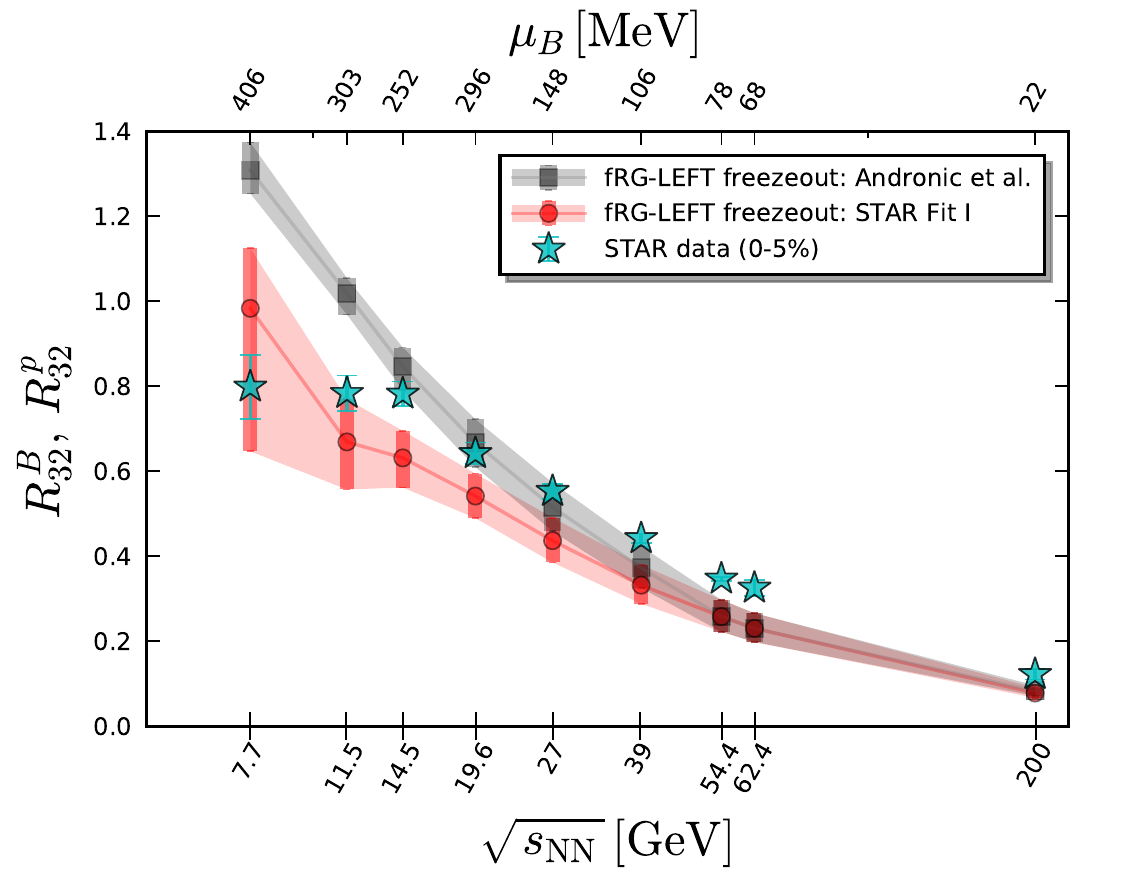}
\includegraphics[width=0.48\textwidth]{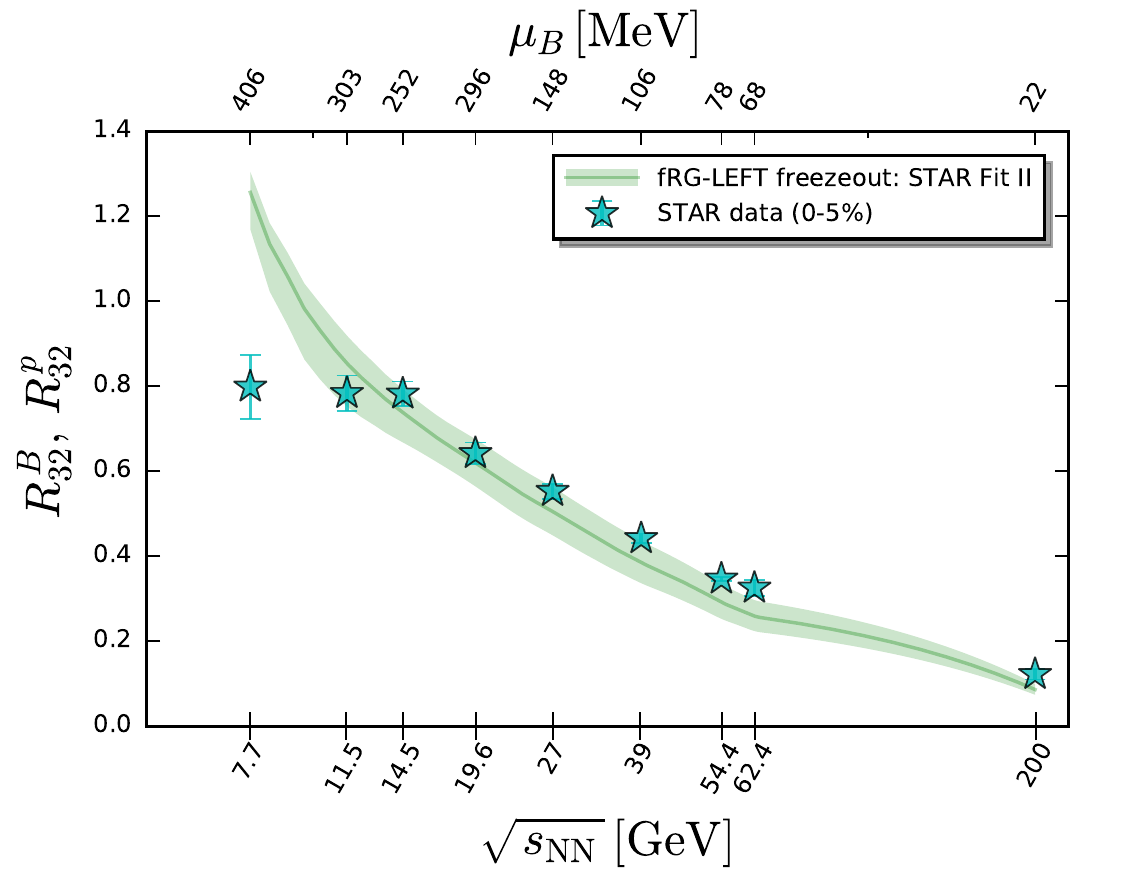}
\caption{Baryon number fluctuation $R^{B}_{32}$ as a function of the collision energy in comparison to STAR-data for $R^{p}_{32}$ (0-5\%) centrality \cite{Adam:2020unf}. 
Left panel: the freeze-out points are those from Andronic {\it et al.} \cite{Andronic:2017pug} (grey) and the STAR experiment \cite{Adamczyk:2017iwn} (red).
Right panel: The freeze-out curve, STAR Fit II, is obtained from the freeze-out parameters of the STAR experiment \cite{Adamczyk:2017iwn}. 
The theoretical error bands show a highly correlated error, and should be interpreted as a family of curves with the same qualitative behaviour as the central curve. For more explanations see \sec{subsec:freezeoutCurve} with \fig{fig:phasediagram}. }\label{fig:R32-sqrtS}
\end{figure*}
%

%
\begin{figure}[t]
\includegraphics[width=0.45\textwidth]{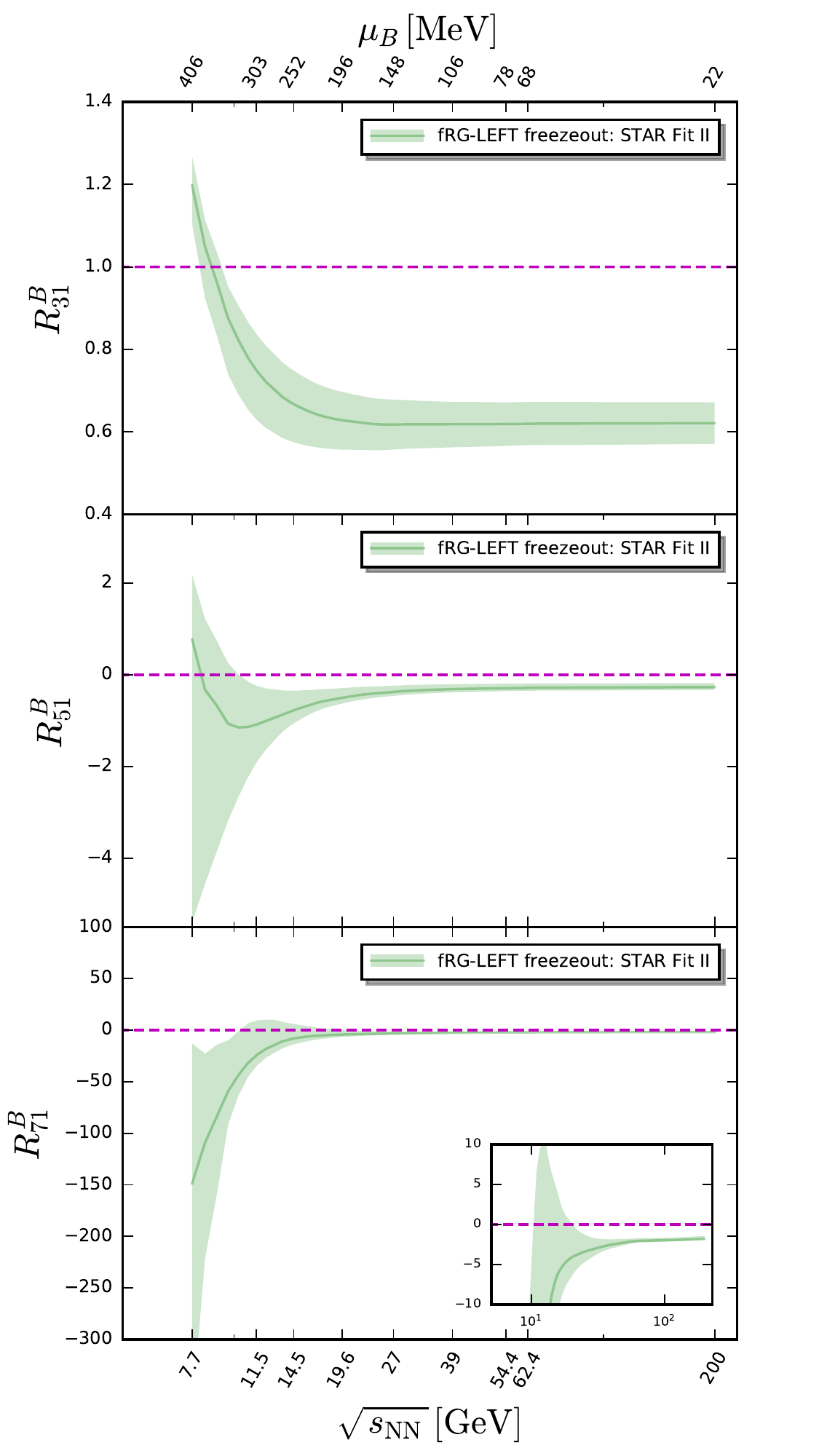}
\caption{Baryon number fluctuations $R^{B}_{31}$ (top), $R^{B}_{51}$ (middle), and $R^{B}_{71}$ (bottom) as functions of the collision energy. 
The freeze-out curve, STAR Fit II, is obtained from the freeze-out parameters of the STAR experiment \cite{Adamczyk:2017iwn}. 
The theoretical error bands show a highly correlated error, and should be interpreted as a family of curves with the same qualitative behaviour as the central curve. For more explanations see \sec{subsec:freezeoutCurve} with \fig{fig:phasediagram}. }\label{fig:Rm1-sqrtS}
\end{figure}
%
The analysis in the previous sections entails that in the present QCD-assisted LEFT the non-monotonic behaviour of  baryon-number fluctuations is triggered by the sharpening of the chiral crossover. This is highly non-trivial, since it is evident, e.g., from \Fig{fig:R42R62R82-T-muB0to400} that neither a system only in the hadronic- nor only in the QGP phase could produce the beam-energy dependencies shown in  \fig{fig:Rm2-sqrtS}. 

In conclusion, the agreement of $R^B_{42}$, computed in the QCD-assisted LEFT and the measured beam-energy dependence of $R^p_{42}$ shows, that the latter could be a signature for the presence of a sharpening crossover between these two phases. Whether or not it also signals the onset of the critical region will be subject of a future improved study. In the context of this latter study we also emphasise that the non-universal properties of the LEFT such as the existence and location of the CEP may not quantitatively agree with QCD as the latter regime lies outside the LEFT-regime with quantitative reliability. Still, the present LEFT probably has the same qualitative non-universal properties at large chemical potential, and it certainly has the same universal ones. 

A non-monotonic energy dependence for the fluctuations is a highly relevant experimental observation, since this behaviour has been proposed as an experimental signature of a CEP \cite{Stephanov:1999zu, Stephanov:2011pb}. The present analysis based on QCD-assisted LEFT model demonstrates that the non-monotonic behaviour of fluctuations can serve as an \textit{indication} of a CEP, but is not necessarily a smoking gun signature for it. The latter requires the extraction of critical scaling, or similar definite signatures such as the detection of a first order regime for large $\mu_B$, etc.\,. 

Still, the non-monotonic behaviour observed in both theory and experiments is a clear \textit{signature} for interesting strongly correlated physics, whose uncovering requires joint and intensified effort of both, theory and experiment. Of course, whether or not these properties carry over completely to QCD remains to be seen. 

Note also that the non-monotonic regime is far away from that covered by a simple extrapolation of the Taylor expansion at $\mu_B=0$. It might be covered by a resummation of the latter, which can already be investigated within the present QCD-assisted LEFT. Constraints on such a resummation should also make use of odd hyper-order fluctuations at finite chemical potential, that are readily computed in the present setup: \\[-1ex] 

A prominent and relevant example is $R_{32}$, already measured in the STAR experiment. In \fig{fig:R32-sqrtS} we show our predictions for $R^B_{32}$ computed in the current QCD-assisted LEFT on the different freeze-out curves defined in \sec{subsec:freezeoutCurve}, see in particular \fig{fig:phasediagram}. In this section it also has been argued, that our best-informed freeze-out curve is given by STAR Fit II. The respective results are shown in the right panel of \fig{fig:R32-sqrtS} in comparison with the STAR data for $R^p_{32}$ (0-5\% centrality). Indeed, these results show the best compatibility with the experimental data. Moreover, within the respective systematic and statistical errors the theoretical results with the freeze-out curve STAR Fit II and the experimental data agree down to collision energies of $\sqrt{s_\mathrm{NN}}\approx$ 14.5\, GeV or $\mu_B\approx 250$\,MeV. 

Interestingly, below $\sqrt{s_\mathrm{NN}}\approx$ 14.5\,GeV the experimental data show a plateau, which is not present in the theoretical prediction. While this is in the large chemical potential regime, in which the LEFT gradually looses its predictive power, also the respective functional first principles QCD computation in \cite{Fischer:2018sdj, Fu:2019hdw, Gao:2020qsj, Gao:2020fbl}, based on a grand potential, do not show any sign of new physics in this regime. This suggests that for  $\sqrt{s_\mathrm{NN}}\lesssim$ 14.5\, GeV at least one of the implicit assumptions underlying the  identification of $R^B_{nm}$ with $R^p_{nm}$ within a grand canonical ensemble with variable baryon charge (density) for given beam energies breaks down. As discussed before, this asks for a re-assessment of the identification of baryon and proton number fluctuations, finite volume effects and finite volume fluctuations, the determination of the freeze-out curve for smaller collision energies, the evaluation of non-equilibrium effects such as transport, and finally the use of the grand potential in the theory computations. While highly relevant and interesting, this goes far beyond the scope of the present work and we defer any further investigation to future work. 

The above example of $R_{32}$ demonstrates very impressively, that the odd (hyper-order) fluctuation observables encode highly relevant physics information which may be difficult or even impossible to extract from the even orders. As a first step in this direction, finally aiming at a resummation of the $\mu_B$-expansion that allows us to go beyond the validity regime of the Taylor expansion, we also have computed the fluctuation observables $R_{31}, R_{51}, R_{71}$ on the freeze-out curve STAR Fit II in \fig{fig:Rm1-sqrtS}. An experimental confirmation of the respective predictions at least for the lower orders would be highly desirable. 

The discussion in this section leaves us with the highly exciting possibility of unravelling the location and properties of a potential CEP within a combined experiment-theory analysis: First principle QCD at finite density should provide us with a prediction for the location of the CEP in terms of hyper-order fluctuations,  $\textrm{Loc}_\textrm{\tiny{CEP}}(R_{nm})$. This would allow us to use the experimental data on hyper-order fluctuation observables $R_{nm}^p$ as input. We emphasise that this prediction does not necessitate the observation of critical behaviour in the $R_{nm}$, but utilises the details of the non-monotonicity of the $R_{nm}$. 

In summary, such an analysis does explicitly not rely on the \textit{universal} property of critical scaling measured in the $R_{nm}$. Indeed, it uses the \textit{non-universal} properties of the $R_{nm}$ to predict the \textit{non-universal} location of the CEP, and hence is far more robust. We hope to report on this matter in the near future. 


\section{Conclusions}
\label{sec:summary}

In this work we have computed baryon number fluctuations up to tenth order with a QCD-assisted low-energy effective theory. This LEFT incorporates quantum, thermal and density fluctuations from momentum scales less than 700\,MeV within the functional renormalisation group approach, and is embedded in QCD, for details see \sec{sec:FRG}. The quantitative predictability has been benchmarked with a comparison of baryon number fluctuations at $\mu_B=0$ up to the eighth order from the lattice, see \sec{subsec:hyper-order0}. Our results are in quantitative agreement with that from the Wuppertal-Budapest collaboration, and are compatible with that of the HotQCD collaboration, as shown in \fig{fig:R42R62R82-T-muB0}. 

Our direct computation at finite $\mu_B$, presented in \sec{subsec:hyper-ordermuB}, has allowed us to assess the range of validity of the Taylor expansion of the free energy of QCD around $\mu_B = 0$. Such an expansion is commonly used to extrapolate lattice results to finite density. We have shown that the expansion up to tenth order in $\mu_B/T$ is only valid for $\mu_B/T \lesssim 1.5$ in the chiral crossover regime, see \fig{fig:R42R62expansion-muBoT}. Beyond this range, the Taylor expansion, at least to this order, fails to even capture the qualitative behaviour of the fourth- and sixth-order baryon number fluctuations. Thus, results for fluctuations at the freeze-out curve based on a Taylor expansion around $\mu_B=0$ should be interpreted with great caution for $\mu_B/T \gtrsim 1.5$, as relevant physical effects might not be captured by this extrapolation.

The main goal of the current work was the computation of baryon number fluctuations and in particular hyper-order fluctuations along the freeze-out curve at collision energies $\sqrt{s}\gtrsim 7.7$\,GeV. The respective results are discussed in \sec{subsec:freezout}. They have been compared to experimental data of net-proton number cumulants from STAR for different estimates for the freeze-out curve, see \fig{fig:Rm2-sqrtS}. 
Our result for the kurtosis, $R^B_{42}$, is in good agreement with the experimental data for collision energies $\sqrt{s}\gtrsim 7.7$\,GeV. In particular the increasing trend at lower beam energies $\sqrt{s}\lesssim 19.6$\,GeV is captured well. This non-monotonicity is also present in the hyper-order fluctuations $R^B_{62}, R^B_{82}$. We also note that a comprehensive comparison for the higher order cumulants is not possible due to the lack of experimental data. Accordingly, our results in \fig{fig:Rm2-sqrtS} are predictions that await experimental verification. 

We have also investigated the twofold origin of the non-monotonicity for $\sqrt{s}\lesssim 19.6$\,GeV in the present LEFT: First, for increasing chemical potential the chiral crossover gets sharper. Secondly, for smaller beam energies the freeze-out temperature may move away from the pseudo-critical temperatures. In the current setup both phenomena happen far away from a potential critical end point in the LEFT located at $\sqrt{s}_\textrm{CEP} \lesssim 3$\,GeV. The latter regime is also safely outside the reliability regime of the current setup, which gradually looses reliability for $\sqrt{s}\lesssim 27$\,GeV. However, its qualitative features may well be present in QCD. The current LEFT-results and its upgrades towards first principle QCD can be compared with the future experimental results of the high statistic data taken from the second phase of RHIC beam energy scan (BES-II, 2019-2021). From the year of 2018 to 2020, the STAR experiment has collected high statistics data of Au+Au collisions at $\sqrt{s_{\mathrm{NN}}}$ = 9.2, 11.5, 14.6, 19.6 and 27 GeV in the collider mode, and $\sqrt{s_{\mathrm{NN}}}$ = 3.0 -- 7.7 GeV in the fixed target mode. These data give us access to the QCD phase structure for baryon chemical potential up to $\mu_{B}$ $\approx$ 720 MeV. 

Related further steps in a comprehensive understanding of the physics of fluctuations in a heavy ion collision has been undertaken in \sec{sec:CEP}, where we have presented results for odd fluctuations observables $R^B_{31},  R^B_{51}, R^B_{71}$ as well as $R^B_{32}$. While the former observables have not been measured yet, $R^B_{32}$ agrees quantitatively within the systematic and statistical error with the STAR measurement for $R^p_{32}$ for collision energies $\sqrt{s_\mathrm{NN}}\gtrsim$ 14.5\, GeV. For smaller energies, the experimental data show a plateau, that may indicate the loss of one or several of the underlying assumption in the identification of theoretical equilibrium computations of baryon number fluctuations $R^B_{nm}$ in a grand canonical ensemble with the experimental results for proton number fluctuations on the freeze-out curve. 

In summary, the non-monotonicities of hyper-order fluctuations, observed both in experiment and theory, are important  \textit{signatures} for interesting physics in the border regime between quark-gluon plasma and the hadron phase. This of course can include a potential CEP, and in any case deserves further investigation from both experiment and theory. In particular, we envisage that experimental data of fluctuation observables and their dependence on collision energy allow us to constrain the onset regime of this strongly correlated physics/CEP. Importantly, such a prediction does not rely on the observation of critical scaling in the hyper-order fluctuations, but is far more robust, for more details see  \sec{sec:CEP}. We hope to report on this in the near future.


\begin{acknowledgments}
We thank Jens Braun, Rob Pisarski, Bernd-Jochen Schaefer and Nu Xu for discussions. The work was supported by the National Key Research and Development Program of China (Grant No. 2020YFE0202002 and 2018YFE0205201) and the National Natural Science Foundation of China under (Grant No. 11775041, 11828501, 11890711 and 11861131009). The work is also supported by EMMI, and the BMBF grant 05P18VHFCA. It is part of and supported by the DFG Collaborative Research Centre SFB 1225 (ISOQUANT) and the DFG under Germany’s Excellence Strategy EXC - 2181/1 - 390900948 (the Heidelberg Excellence Cluster STRUCTURES).

\end{acknowledgments}

	
\appendix


\section{The fRG-approach to QCD \& LEFTs}\label{app:fRG}
	
The functional renormalisation group or flow equation for QCD provides the evolution of its effective action $\Gamma_k$ with an  infrared cutoff scale $k$. Here we use  the setup with \textit{dynamical hadronisation}, \cite{Gies:2001nw, Gies:2002hq, Pawlowski:2005xe, Braun:2008pi, Floerchinger:2009uf, Fu:2019hdw}. The formulation used here has been developed in \cite{Mitter:2014wpa, Braun:2014ata, Rennecke:2015eba, Cyrol:2017ewj, Fu:2019hdw}. Its current form has been described and further developed in \cite{Fu:2019hdw}, and for further details we refer to this work. The flow equation of the QCD effective action reads, 
\begin{align}
\partial_t\Gamma_k[\Phi]=&\frac{1}{2}\mathrm{Tr}\Big(G_{AA,k}\partial_t R_{A,k}\Big)-\mathrm{Tr}\Big(G_{c\bar c,k}\partial_t R_{c,k}\Big)\nonumber\\[2ex]
&-\mathrm{Tr}\Big(G_{q\bar q,k}\partial_t R_{q,k}\Big)+\frac{1}{2}\mathrm{Tr}\Big(G_{\phi\phi,k}\partial_t R_{\phi,k}\Big)\,.\label{eq:QCDflow}
\end{align}
In \eq{eq:QCDflow}, the $\Phi=(A, c, \bar c, q,\bar q,\phi)$ is a superfield that comprises all fields. This also includes hadronic (composite) low energy degrees of freedom introduced by dynamical hadronisation. The $G$'s and $R$'s are the propagators and regulators of the different fields, respectively. Diagrammatically it is depicted in \Fig{fig:QCD_equation}. For more works on QCD-flows at finite temperature and density see \cite{Braun:2007bx, Braun:2008pi, Braun:2009gm, Mitter:2014wpa,  Braun:2014ata, Rennecke:2015eba,  Cyrol:2016tym, Cyrol:2017ewj, Cyrol:2017qkl, Fu:2019hdw, Braun:2020ada, Braun:2020mhk}, for reviews on QCD and LEFTs for QCD see \cite{Litim:1998nf, Berges:2000ew, Pawlowski:2005xe, Schaefer:2006sr, Gies:2006wv, Rosten:2010vm, Braun:2011pp, Pawlowski:2014aha, Dupuis:2020fhh}.
	
For scales $k\lesssim 1$\,GeV, the gluon decouples from the system due to its confinement-related mass gap. For these momentum scales, the (off-shell) dynamics of QCD is dominated by quarks and the emergent composite hadronic degrees of freedom. In particular, the lowest lying meson multiplet, and specifically the $\pi$ meson is driving the dynamics. The pion is the pseudo-Goldstone boson of strong chiral symmetry breaking, and hence is the lightest hadron with a mass $\sim 140$ MeV in the vacuum. 

Consequently, in this regime with $k\lesssim 1$\,GeV, the flow equation of the QCD effective action in \Eq{eq:QCDflow} is reduced to, 
\begin{align}
\partial_t\Gamma_k[\Phi]=&-\mathrm{Tr}\Big(G_{q\bar q,k}\partial_t R_{q,k}\Big)+\frac{1}{2}\mathrm{Tr}\Big(G_{\phi\phi,k}\partial_t R_{\phi,k}\Big)\,,\label{eq:LEFTflow}
\end{align}
where $R_{q,k}$ and $R_{\phi,k}$ are the regulators for the quark and meson fields, respectively. The full propagators in \eq{eq:LEFTflow} read, 
\begin{align}
G_{q\bar q/\phi\phi,k}=&\left(\frac{1}{\Gamma^{(2)}_k[\Phi]+R_k}\right)_{q\bar q/\phi\phi}\,,\label{eq:Gfull}
\end{align}
with $\Gamma^{(2)}_k[\Phi]=\delta^2\Gamma_k[\Phi]/\delta \Phi^2$. In this work we employ  $3d$-flat or Litim regulators \cite{Litim:2000ci, Litim:2001up, Litim:2006ag}, 
\begin{align}\nonumber 
R_{\phi,k}(q_0,\bm{q})&=Z_{\phi,k}\bm{q}^2 r_B(\bm{q}^2/k^2)\,, \\[2ex] 
R_{q,k}(q_0,\bm{q})&=Z_{q,k}i\bm{\gamma} \cdot \bm{q} r_F(\bm{q}^2/k^2)\,, \label{eq:Rk}
\end{align} 
with 
\begin{align}\nonumber 
r_B(x)&=\left( \frac{1}{x}-1 \right)\Theta(1-x)\,,\\[2ex] 
r_F(x)&=\left( \frac{1}{\sqrt{x}}-1 \right)\Theta(1-x)\,,  \label{eq:rk}
\end{align} 
where $\Theta(x)$ denotes the Heaviside step function. Inserting the effective action \eq{eq:action} into the flow equation \eq{eq:LEFTflow}, we arrive at 
\begin{align}
\partial_t V_{\mathrm{mat},k}(\rho)=&\frac{k^4}{4\pi^2} \bigg [\big(N^2_f-1\big) l^{(B,4)}_{0}(\tilde{m}^{2}_{\pi,k},\eta_{\phi,k};T)\nonumber\\[2ex]
&+l^{(B,4)}_{0}(\tilde{m}^{2}_{\sigma,k},\eta_{\phi,k};T)\nonumber\\[2ex]
&-4N_c N_f l^{(F,4)}_{0}(\tilde{m}^{2}_{q,k},\eta_{q,k};T,\mu)\bigg]\,, \label{eq:flowV}
\end{align}
where the threshold functions $l^{(B/F,4)}_{0}$ as well as other threshold functions used in the following can be found in e.g., \cite{Fu:2019hdw,Yin:2019ebz}. The dimensionless renormalised quark and meson masses read, 
\begin{align}
\tilde{m}^{2}_{q,k}=&\frac{h^{2}_{k}\rho}{2k^2Z^{2}_{q,k}}\,, \qquad \tilde{m}^{2}_{\pi,k}=\frac{V'_{\mathrm{mat},k}(\rho)}{k^2 Z_{\phi,k}}\,, \\[2ex]
\tilde{m}^{2}_{\sigma,k}=&\frac{V'_{\mathrm{mat},k}(\rho)+2\rho V''_{\mathrm{mat},k}(\rho)}{k^2 Z_{\phi,k}}\,.\label{eq:mqk}
\end{align}
The anomalous dimensions for the quark and meson fields in \Eq{eq:flowV} are defined as
\begin{align}
\eta_{q,k}&=-\frac{\partial_t Z_{q,k}}{Z_{q,k}}\,,\qquad
\eta_{\phi,k}=-\frac{\partial_t Z_{\phi,k}}{Z_{\phi,k}}\,,
\end{align}
respectively. Accordingly, the flow equation for the mesonic anomalous dimension is obtained from the (spatial) momentum derivative w.r.t. $\bm{p}^2$ of the pion two-point function, to wit, 
\begin{align}
\eta_{\phi,k}&=-\frac{1}{3Z_{\phi,k}}\delta_{ij}\frac{\partial}{\partial \bm{p}^2}\frac{\delta^2 \partial_t \Gamma_k}{\delta \pi_i(-p) \delta \pi_j(p)}\Bigg|_{\substack{p_0=0\\ \bm{p}=0}}\,.\label{eq:etaphi}
\end{align}
The approximation \eq{eq:action} to the effective action together with \eq{eq:etaphi} are based on two approximations: Firstly, in \Eq{eq:action} we have dropped the field-dependence of $Z_\phi$, which would lead to different $Z_\pi$ and $Z_\sigma$. In \eq{eq:etaphi} we have identified $Z_\phi=Z_\pi$, and hence also $Z_\sigma=Z_\pi$. This is motivated by the fact that the meson dynamics are only dominant in the broken regime where the three pions are far lighter than the single sigma mode, which quickly decouples. Hence, the three pions drive the dynamics. 

Furthermore, in \Eq{eq:action} we do not distinguish between spatial and temporal components of $Z_\phi$. For finite temperature and density, the Euclidean $\mathrm{O}(4)$ rotation symmetry is broken, as the heat bath of density singles out a rest frame. This entails, that $\eta_{\phi,k}$ splits into $\eta_{\phi,k}^{\perp}$ and $\eta_{\phi,k}^{\parallel}$, the components transverse and longitudinal to the heat bath/density. We have used the approximation $\eta_{\phi,k}=\eta_{\phi,k}^{\perp}$ as we have three spatial directions. The influence of the splitting of $\eta_{\phi,k}$ on the thermodynamics and baryon number fluctuations has been investigated in detail e.g.\ in \cite{Yin:2019ebz}. There it has been found that the impact is small, supporting the reliability of the present approximation. 

Similarly, the quark anomalous dimension is obtained by projecting the relevant flow onto the vector channel of the 1PI quark--anti-quark correlation function, 
\begin{align}
\eta_{q}=&\frac{1}{4 Z_{q,k}} \nonumber\\[1ex]
&\hspace{-.8cm}\times \mathrm{Re}\left[\frac{\partial}{\partial \bm{p}^2}\mathrm{tr}
\left(i \bm{\gamma}\cdot\bm{p}\left(-\frac{\delta^2}{\delta\bar{q}(p)
\delta q(p)}\partial_t \Gamma_k\right)\right)\right]\Bigg|_{\substack{p_{0,ex}\\ \bm{p}=0}}\,.   \label{eq:etapsi}
\end{align}
In \eq{eq:etapsi}, the spatial momentum is set to zero, ${\bm p}=0$ as in the mesonic case: vanishing momentum is most relevant to the flow of effective potential in \Eq{eq:flowV}. Note, that the lowest fermionic Matsubara frequency is non-vanishing. We use $p_{0,ex}\neq 0$, its value is further described in \app{app:flowV}, based on \cite{Fu:2015naa, Fu:2016tey}. 

As is implicit in \Eq{eq:etapsi}, the flow of the quark two-point function is complex-valued at non-vanishing chemical potential. This originates in the Silver-Blaze property of QCD at $T=0$. For quark correlation functions this entails that they are functions of $p_0-i \mu_q$ already before the onset of the baryon density, for a discussion in the present fRG-approach see \cite{Khan:2015puu, Fu:2015naa, Fu:2016tey}. In turn, the couplings are still real (i.e.\ real functions of the  complex variable $p_0-i \mu_q$) in particular below the density onset. Hence, couplings (i.e.\ expansion coefficients in a Taylor expansion in momenta) are real. This is readily seen in a resummation of the external frequency of the quark propagator \cite{Fu:2016tey}. Without resummation they are obtained from a projection on the real part of the flow, see \eq{eq:etapsi}. 

This projection is also used for the Yukawa coupling. Within the present approximation, the flow equation of the (real) Yukawa coupling is given by, 
\begin{align}
\partial_t h_k&=\frac{1}{2 \sigma}\mathrm{Re}\left[\mathrm{tr}\left(-\frac{\delta^2}{\delta\bar{q}(p)
\delta q(p)}\partial_t \Gamma_k\right)\right]\Bigg|_{\substack{p_{0,ex}\\ \bm{p}=0}}\,.  \label{eq:dth}
\end{align}
The explicit expressions for the meson and quark anomalous dimensions, as well as the flow of the Yukawa coupling can be found in \app{app:flowV}.


\section{Flow equations for $V_k(\rho)$, $h_k$, and $\eta_{\phi,q}$}\label{app:flowV}
	
The flow equation for the effective potential is given in \Eq{eq:flowV}. To resolve its field dependence, we use a Taylor expansion about a $k$-dependent $\rho$-value $\kappa_k$,  
\begin{align}
V_{\mathrm{mat}, k}(\rho)&=\sum_{n=0}^{N_v}\frac{\lambda_{n,k}}{n!}(\rho-\kappa_k)^n\,, \label{eq:VTaylor}
\end{align}
with the running expansion coefficients $\lambda_{n,k}$. Here, $N_v$ is the maximal order of Taylor expansion included in the numerics. $N_v=5$ is adopted in this work, which is large enough to guarantee the convergence of expansion, for more details, see e.g., \cite{Pawlowski:2014zaa,Yin:2019ebz}. It is more convenient to rewrite \eq{eq:VTaylor} by means of the renormalised variables, i.e.,
\begin{align}
\bar V_{\mathrm{mat}, k}(\bar \rho)&=\sum_{n=0}^{N_v}\frac{\bar\lambda_{n,k}}{n!}(\bar \rho-\bar \kappa_k)^n\,,\label{eq:VbarTaylor}
\end{align}
with $\bar V_{\mathrm{mat}, k}(\bar \rho)=V_{\mathrm{mat}, k}(\rho)$, $\bar \rho=Z_{\phi,k} \rho$, $\bar \kappa_k=Z_{\phi,k}\kappa_k$, and $\bar \lambda_{n,k}=\lambda_{n,k}/(Z_{\phi,k})^n$. Inserting \eq{eq:VbarTaylor} into the l.h.s.\ of \Eq{eq:flowV} leads us to, 
\begin{align}
&\partial^n_{\bar \rho}\left(\partial_t\big|_{\rho} \bar V_{\mathrm{mat}, k}(\bar \rho)\right)\Big|_{\bar \rho=\bar \kappa_k}\nonumber\\[2ex]
=&(\partial_t -n\eta_{\phi,k})\bar{\lambda}_{n,k}-(\partial_t \bar \kappa_k+\eta_{\phi,k}\bar \kappa_k)\bar \lambda_{n+1,k}\,.\label{eq:drhoV}
\end{align}
In the present work, we use the EoM of $\rho$ as our expansion point. With \eq{eq:action} this yields, 
\begin{align}
\frac{\partial}{\partial \bar \rho}\Big(\bar V_{\mathrm{mat}, k}(\bar \rho)-\bar c_k
\bar \sigma \Big)\bigg \vert_{\bar\rho=\bar \kappa_k}&=0\,, \label{eq:Vstat}
\end{align}
with $ \bar \sigma=Z_{\phi,k}^{1/2} \sigma$ and $\bar c_k=Z_{\phi,k}^{-1/2} c$, with a cutoff-independent  $c$. Another commonly used expansion point is a fixed expansion point, $\partial_t \kappa_k=0$. For further details on these two different expansion approaches, and their respective convergence properties see   \cite{Pawlowski:2014zaa, Braun:2014ata, Rennecke:2015lur, Fu:2015naa, Rennecke:2016tkm, Yin:2019ebz}. 

From \eq{eq:drhoV} and \eq{eq:Vstat} we get the flow equation for the expansion point, 
\begin{align}
\partial_t \bar \kappa_k=&\,-\frac{\bar c_k^2}{\bar{\lambda}_{1,k}^3+\bar c_k^2\bar{\lambda}_{2,k}}\Bigg[\partial_{\bar \rho}\left(\partial_t\big|_{\rho} \bar V_{\mathrm{mat}, k}(\bar \rho)\right)\Big|_{\bar \rho=\bar \kappa_k}\nonumber \\[1ex]
&\hspace{1.5cm}+\eta_{\phi,k}\left(\frac{\bar{\lambda}_{1,k}}{2}+\bar\kappa_k\bar{\lambda}_{2,k}\right)\Bigg]\,.\label{eq:flowkappa}
\end{align}
The meson anomalous dimension in \Eq{eq:etaphi} reads, 
\begin{align}
\eta_{\phi,k}=&\frac{1}{6\pi^2}\Bigg\{\frac{4}{k^2} \bar{\kappa}_k(\bar{V}''_k(\bar{\kappa}_k))^2\mathcal{BB}_{(2,2)}(\tilde{m}^{2}_{\pi,k},\tilde{m}^{2}_{\sigma,k};T)\nonumber\\[1ex]
&\hspace{.6cm}+N_c\bar{h}^{2}_{k}\bigg[\mathcal{F}_{(2)}(\tilde{m}^{2}_{q,k};T,\mu)(2\eta_{q,k}-3)\nonumber\\[1ex]
&\hspace{.6cm}-4(\eta_{q,k}-2)\mathcal{F}_{(3)}(\tilde{m}^2_{q,k};T,\mu)\bigg]\Bigg\}\,, \label{eq:etaphi2}  
\end{align} 
The quark anomalous dimension in \Eq{eq:etapsi} reads, %
\begin{align}
\eta_{q,k}=&\frac{1}{24\pi^2N_f}(4-\eta_{\phi,k})\bar{h}^{2}_{k}\nonumber\\[2ex]
&\times\bigg\{ (N^{2}_{f}-1)\mathcal{FB}_{(1,2)}(\tilde{m}^{2}_{q,k},\tilde{m}^{2}_{\pi,k};T,\mu,p_{0,ex})\nonumber\\[2ex]
&+\mathcal{FB}_{(1,2)}(\tilde{m}^{2}_{q,k},\tilde{m}^{2}_{\sigma,k};T,\mu,p_{0,ex})\bigg\}\,.  \label{eq:etapsi2}
\end{align} 
In the threshold function $\mathcal{FB}$'s we have employed $p_{0,ex}=\pi T$ for the finite temperature sector and $p_{0,ex}=\pi T\exp\{-k/(\pi T)\}$ for the vacuum sector. 
This choice guarantees a consistent temperature dependence for all $k$, which is particularly relevant for the thermodynamics in the low temperature region \cite{Fu:2015naa}. This can be resolved by means of a full frequency summation of the quark external leg \cite{Fu:2016tey}, and the present procedure mimics this physical behaviour. 

The flow of the Yukawa coupling in \Eq{eq:dth} reads, 
\begin{align}
\partial_t\bar{h}_k=&\left(\frac{1}{2}\eta_{\phi,k}+\eta_{q,k}\right)\bar{h}_k(\bar{\rho})\nonumber\\[2ex]
&\hspace{-.8cm}+\frac{\bar{h}^3_k}{4\pi^2N_f}\bigg[L^{(4)}_{(1,1)}(\tilde{m}^{2}_{q,k},\tilde{m}^{2}_{\sigma,k},\eta_{q,k},\eta_{\phi,k};T,\mu,p_{0,ex})\nonumber\\[1ex]
&\hspace{-.8cm}-(N^{2}_{f}-1)L^{(4)}_{(1,1)}(\tilde{m}^{2}_{q,k},\tilde{m}^{2}_{\pi,k},\eta_{q,k},\eta_{\phi,k};T,\mu,p_{0,ex})\bigg]\,.\label{eq:dth2}  
\end{align} 
Explicit expressions of all the threshold functions mentioned above, such as $\mathcal{BB}$, $\mathcal{F}$'s, $\mathcal{FB}$'s, and $L$ can be found in e.g., \cite{Fu:2019hdw,Yin:2019ebz}. 
	
In summary, the flow equations \eq{eq:flowV}, \eq{eq:drhoV}, \eq{eq:flowkappa}, \eq{eq:dth2}, supplemented with \eq{eq:etaphi2} and \eq{eq:etapsi2}, constitute a closed set of ordinary differential equations, which is evolved from the UV cutoff $k=\Lambda$ to the IR limit $k=0$. 


\section{Initial conditions}
\label{app:Ini}

To solve the flow equation, we need to specify initial conditions. To this end, we choose initial values at a scale $k=\Lambda$ such that known observables of QCD in the vacuum at $k=0$, such as the pion mass and decay constant, are reproduced. The effective potential at the UV cutoff reads, 
\begin{align}
V_{\mathrm{mat},  \Lambda}(\rho)=\frac{\lambda_{\Lambda}}{2}\rho^2+\nu_{\Lambda}\rho \,. 
\end{align}
We initialise the flows at $\Lambda=700$\,MeV. The input parameters of the QCD-assisted LEFT are given in \tab{tab:InitialValues}. 

\begin{table}[t]
	\begin{center}
		\begin{tabular}{|r ||c|| c |}
			\hline & & \\[-1ex]
			Observables & Value & Parameter in $\Gamma_\Lambda$, \  $\Lambda=700\,$MeV\\[1ex]
			\hline& & \\[-1ex]
\begin{tabular}{r} $m_{\pi,\textrm{pol}}$ [MeV]\\[1ex] 
$m_\sigma$ [MeV]  \\[2ex]
$f_{\pi}$ [MeV]
\end{tabular}		      
& 
\begin{tabular}{c} 136\\[1ex]
479\\[2ex]
92 
\end{tabular} 
&  
\begin{tabular}{rcl} 		 $\lambda_{\Lambda}$&=&11\\[1ex]   $\nu_{k=\Lambda}$&=& $(0.830\,\mathrm{GeV})^2$ \\[2ex]
$c_\sigma$ &=& 2.82$\times 10^{-3}\,\text{GeV}^3$  
\end{tabular}\\[5ex]
\hline& &\\[-2ex]
\hline & & \\[-1ex]
\begin{tabular}{r}$\bar m_l$ [MeV] \end{tabular}     &  
300 & \begin{tabular}{rcl} 	\hspace{-1.4cm} $h_\Lambda$ &=& 10.18  \end{tabular}\\[1ex] 
\hline
\end{tabular}
\caption{Observables and related initial values for the LEFT couplings at the initial cutoff scale $\Lambda=700$\,MeV. The parameters are fixed with the pion pole mass $m_{\pi,\textrm{pol}}$, the mass of the sigma resonance, $m_{\sigma}$, and the pion decay constant, $f_\pi\approx  \bar \sigma_\textrm{EoM}$, the expectation value of the sigma-field. The input parameters are those of the initial effective potential $V_\Lambda$: the meson self-coupling $\lambda_\Lambda$ and the meson mass parameter $\nu_\Lambda$. The pion mass is tuned by $c_\sigma$, the parameter of explicit chiral symmetry breaking. Finally, the constituent quark mass is fixed via the initial value for the Yukawa coupling, $h_\Lambda$.}
		\label{tab:InitialValues}
	\end{center}
\end{table}
%


\section{Glue potential}\label{app:gluepot}

The dynamics of the glue sector in QCD is partly imprinted in the glue potential $V_{\mathrm{glue},k}(A_0)$, see \eq{eq:Vtotal}. This has been discussed in \sec{sec:FRG}. This potential is only needed for the determination of the expectation value of the Polyakov loop. Its inherent glue correlation functions are gapped and their backcoupling is suppressed for $k\lesssim 1$\,GeV. 
Accordingly, we can simply drop the scale dependence of the glue potential for the present purposes. This leads us to,  
\begin{align}
V_\mathrm{glue}(L,\bar{L})&=V_{\mathrm{glue},k=0}(A_0)=T^4 \bar V_\mathrm{glue}(L,\bar{L})\,.\label{eq:dimlessVglue}
\end{align}
In \eq{eq:dimlessVglue} we have introduced a dimensionless glue potential $\bar V_\mathrm{glue}$. Its dependence on the temporal gluon background field, $A_0$, is encoded in the traced Polyakov loop $L[A_0]$ and its conjugate $\bar{L}[A_0]$, 
\begin{align}
L(\bm{x})&=\frac{1}{N_c} \left\langle \Tr\, {\cal P}(\bm x)\right\rangle\,,\quad  \bar L (\bm{x})=\frac{1}{N_c} \langle \Tr\,{\cal P}^{\dagger}(\bm x)\rangle \,,\label{eq:Lloop}
\end{align}
with 
\begin{align}
{\cal P}(\bm x)&=\mathcal{P}\exp\Big(ig\int_0^{\beta}d\tau \hat A_0(\bm{x},\tau)\Big)\,, \label{eq:Ploop}
\end{align}
where $\mathcal{P}$ on the r.h.s.\ stands for path ordering. 

%
\begin{table}[b]
\begin{center}
\begin{tabular}{|c||c|c|c|c|c|}
\hline & & & & & \\[-1ex]
& 1 & 2 & 3 & 4 & 5 \\[1ex]
\hline & & & & &  \\[-2ex]
$a_i$ &-44.14& 151.4 & -90.0677 &2.77173 &3.56403 \\[1ex]
\hline & & & & &  \\[-2ex]
$b_i$ &-0.32665 &-82.9823 &3.0 &5.85559  &              \\[1ex]
\hline & & & & &  \\[-2ex]
$c_i$ &-50.7961 &114.038 &-89.4596 &3.08718 &6.72812 \\[1ex]
\hline & & & & &  \\[-2ex]
$d_i$ & 27.0885 &-56.0859 &71.2225 &2.9715 &6.61433 \\[1ex]
\hline
\end{tabular}
\caption{Values of the parameters for the glue potential in \Eq{eq:xT} and \Eq{eq:bT}.}
\label{tab:gluepotCoeffs}
\end{center}\vspace{-0.5cm}
\end{table}
%

In this work we adopt the parametrisation of the glue potential in \cite{Lo:2013hla}, which reads
\begin{align}
V_\text{glue}(L,\bar{L})=& -\frac{a(T)}{2} \bar L L + b(T)\ln M_H(L,\bar{L})\nonumber \\[2ex]
&+ \frac{c(T)}{2} (L^3+\bar L^3) + d(T) (\bar{L} L)^2\,,
\label{eq:polpot}
\end{align}
with the $\mathrm{SU}(N_c)$ Haar measure
\begin{align}
M_H (L, \bar{L})&= 1 -6 \bar{L}L + 4 (L^3+\bar{L}^3) - 3  (\bar{L}L)^2\,.
\end{align}
Both, the parametrisation of glue potential in \Eq{eq:polpot}, as well as determination of relevant parameters in \Tab{tab:gluepotCoeffs}, is based on lattice results of  $\mathrm{SU}(3)$ Yang-Mills theory at finite temperature. This potential does not only reproduce the lattice expectation value of the Polyakov loop and the pressure, but also the correct quadratic fluctuations of the Polyakov loop, \cite{Lo:2013hla}. These fluctuations, and higher ones, are important for the fluctuation observables discussed here \cite{Fu:2015naa, Fu:2016tey}. The coefficients in \Eq{eq:polpot} are temperature-dependent, 
\begin{align}
x(T) &= \frac{x_1 + x_2/(t_r+1) + x_3/(t_r+1)^2}{1 + x_4/(t_r+1) + x_5/(t_r+1)^2}\,,\label{eq:xT}
\end{align}
with $x=a, c, d$, and 
\begin{align}
b(T) &=b_1 (t_r+1)^{-b_4}\left (1 -e^{b_2/(t_r+1)^{b_3}} \right)\,.\label{eq:bT}
\end{align}
In \eq{eq:xT} and \eq{eq:bT} we have used the reduced temperature $t_r=(T-T_c)/T_c$. The parameter values are taken from \cite{Lo:2013hla}, and are collected in \Tab{tab:gluepotCoeffs} for convenience.

The parameters in \tab{tab:gluepotCoeffs} are that of the glue potential in Yang-Mills theory. It has been 
argued and shown in \cite{Pawlowski:2010ht, Haas:2013qwp, Herbst:2013ufa} that unquenching effects in QCD are well captured by a linear rescaling of the reduced temperature in the regime about $T_c$, very similar to the rescaling discussed in \sec{sec:scaleLEFT}. This leads us to, 
\begin{align}
(t_r)_{\text{\tiny{YM}}}&\rightarrow \alpha\,(t_r)_{\text{\tiny{glue}}}\,,\label{eq:Tgluescale}
\end{align}
with
\begin{align}
(t_r)_{\text{\tiny{glue}}}&=
(T-T_c^\text{\tiny{glue}})/T_c^\text{\tiny{glue}}\,.\label{eq:TglueReduced}
\end{align}
In the present work we have used $\alpha=0.75$ and $T_c^\text{\tiny{glue}} =213\, {\rm MeV}$.


	
\bibliography{ref-lib}

\end{document}